\documentclass[conference,10pt,letterpaper,nofonttune]{IEEEtran}
\IEEEoverridecommandlockouts
\usepackage{cite}
\usepackage{listings}
\usepackage{fix-cm}
\usepackage[utf8]{inputenc}
\usepackage{csquotes}
\usepackage{amsmath,amssymb,amsfonts}
\usepackage{algorithmic,algorithm}
\usepackage{textcomp}
\usepackage{gensymb}
\usepackage{xcolor}
\usepackage{subfigure}
\usepackage{comment}
\usepackage{xspace}
\usepackage{balance}
\usepackage{fancyhdr}
\usepackage{newunicodechar}
\usepackage{bm} % Add this in your preamble
\usepackage{siunitx}

\newcommand{\mycomment}[1]{}

\newcommand{\quadriga}{QuaDRiGa\xspace}
\usepackage{soul}
\usepackage{multirow}
\usepackage{booktabs}
\usepackage{tabularx}
\usepackage{tikz}
\usepackage{pgfplots}
\pgfplotsset{compat=1.18}
\usetikzlibrary{plotmarks}
\usepackage{adjustbox}
\pgfplotsset{compat=newest}
\pgfplotsset{plot coordinates/math parser=false}
\newlength\fheight
\newlength\fwidth
\usetikzlibrary{plotmarks,patterns,decorations.pathreplacing,backgrounds,calc,arrows,arrows.meta,spy,matrix,scopes}
\usepgfplotslibrary{patchplots,groupplots}
\usepackage{tikzscale}
\usepackage[draft]{hyperref}

\newif\ifexttikz
\exttikzfalse
\newunicodechar{λ}{$\lambda$}
\ifexttikz
	\usetikzlibrary{external}
	\usepackage{fontspec}
\fi
\usepackage[acronyms,nonumberlist,nopostdot,nomain,nogroupskip,acronymlists={hidden}]{glossaries}
\newglossary[algh]{hidden}{acrh}{acnh}{Hidden Acronyms}
\newacronym{3gpp}{3GPP}{3rd Generation Partnership Project}
\newacronym{4g}{4G}{4th generation}
\newacronym{5g}{5G}{5th generation}
\newacronym{6g}{6G}{6th generation}
\newacronym{5gc}{5GC}{5G Core}
\newacronym{adc}{ADC}{Analog to Digital Converter}
\newacronym{aerpaw}{AERPAW}{Aerial Experimentation and Research Platform for Advanced Wireless}
\newacronym{ai}{AI}{Artificial Intelligence}
\newacronym{aimd}{AIMD}{Additive Increase Multiplicative Decrease}
\newacronym{am}{AM}{Acknowledged Mode}
\newacronym{amc}{AMC}{Adaptive Modulation and Coding}
\newacronym{amf}{AMF}{Access and Mobility Management Function}
\newacronym{sota}{SoTA}{state-of-the-art}
\newacronym{aops}{AOPS}{Adaptive Order Prediction Scheduling}
\newacronym{api}{API}{Application Programming Interface}
\newacronym{xapp}{xApp}{Intelligent Application}
\newacronym{apn}{APN}{Access Point Name}
\newacronym{aqm}{AQM}{Active Queue Management}
\newacronym{ausf}{AUSF}{Authentication Server Function}
\newacronym{avc}{AVC}{Advanced Video Coding}
\newacronym{awgn}{AGWN}{Additive White Gaussian Noise}
\newacronym{balia}{BALIA}{Balanced Link Adaptation Algorithm}
\newacronym{bbu}{BBU}{Base Band Unit}
\newacronym{bdp}{BDP}{Bandwidth-Delay Product}
\newacronym{ber}{BER}{Bit Error Rate}
\newacronym{bf}{BF}{Beamforming}
\newacronym{bler}{BLER}{Block Error Rate}
\newacronym{brr}{BRR}{Bayesian Ridge Regressor}
\newacronym{bs}{BS}{Base Station}
\newacronym{bsr}{BSR}{Buffer Status Report}
\newacronym{bss}{BSS}{Business Support System}
\newacronym{ca}{CA}{Carrier Aggregation}
\newacronym{caas}{CaaS}{Connectivity-as-a-Service}
\newacronym{cb}{CB}{Code Block}
\newacronym{cc}{CC}{Congestion Control}
\newacronym{ccid}{CCID}{Congestion Control ID}
\newacronym{cv2x}{C-V2X}{Cellular Vehicle-to-Everything}
\newacronym{cco}{CC}{Carrier Component}
\newacronym{cdd}{CDD}{Cyclic Delay Diversity}
\newacronym{cdf}{CDF}{Cumulative Distribution Function}
\newacronym{cdn}{CDN}{Content Distribution Network}
\newacronym{cn}{CN}{Core Network}
\newacronym{codel}{CoDel}{Controlled Delay Management}
\newacronym{comac}{COMAC}{Converged Multi-Access and Core}
\newacronym{cord}{CORD}{Central Office Re-architected as a Datacenter}
\newacronym{cornet}{CORNET}{COgnitive Radio NETwork}
\newacronym{cosmos}{COSMOS}{Cloud Enhanced Open Software Defined Mobile Wireless Testbed for City-Scale Deployment}
\newacronym{cots}{COTS}{Commercial Off-the-Shelf}
\newacronym{cp}{CP}{Control Plane}
\newacronym{cpu}{CPU}{Central Processing Unit}
\newacronym{cqi}{CQI}{Channel Quality Information}
\newacronym{cr}{CR}{Cognitive Radio}
\newacronym{cran}{CRAN}{Cloud \gls{ran}}
\newacronym{crs}{CRS}{Cell Reference Signal}
\newacronym{csi}{CSI}{Channel State Information}
\newacronym{csirs}{CSI-RS}{Channel State Information - Reference Signal}
\newacronym{cu}{CU}{Central Unit}
\newacronym{d2tcp}{D$^2$TCP}{Deadline-aware Data center TCP}
\newacronym{d3}{D$^3$}{Deadline-Driven Delivery}
\newacronym{dac}{DAC}{Digital to Analog Converter}
\newacronym{dag}{DAG}{Directed Acyclic Graph}
\newacronym{das}{DAS}{Distributed Antenna System}
\newacronym{dash}{DASH}{Dynamic Adaptive Streaming over HTTP}
\newacronym{dc}{DC}{Dual Connectivity}
\newacronym{dccp}{DCCP}{Datagram Congestion Control Protocol}
\newacronym{dce}{DCE}{Direct Code Execution}
\newacronym{dci}{DCI}{Downlink Control Information}
\newacronym{dctcp}{DCTCP}{Data Center TCP}
\newacronym{dl}{DL}{Downlink}
\newacronym{dmr}{DMR}{Deadline Miss Ratio}
\newacronym{dmrs}{DMRS}{DeModulation Reference Signal}
\newacronym{drlcc}{DRL-CC}{Deep Reinforcement Learning Congestion Control}
\newacronym{drs}{DRS}{Discovery Reference Signal}
\newacronym{du}{DU}{Distributed Unit}
\newacronym{e2e}{E2E}{end-to-end}
\newacronym{ecaas}{ECaaS}{Edge-Cloud-as-a-Service}
\newacronym{ecn}{ECN}{Explicit Congestion Notification}
\newacronym{edf}{EDF}{Earliest Deadline First}
\newacronym{embb}{eMBB}{Enhanced Mobile Broadband}
\newacronym{empower}{EMPOWER}{EMpowering transatlantic PlatfOrms for advanced WirEless Research}
\newacronym{enb}{eNB}{evolved Node Base}
\newacronym{endc}{EN-DC}{E-UTRAN-\gls{nr} \gls{dc}}
\newacronym{epc}{EPC}{Evolved Packet Core}
\newacronym{eps}{EPS}{Evolved Packet System}
\newacronym{es}{ES}{Edge Server}
\newacronym{etsi}{ETSI}{European Telecommunications Standards Institute}
\newacronym[firstplural=Estimated Times of Arrival (ETAs)]{eta}{ETA}{Estimated Time of Arrival}
\newacronym{eutran}{E-UTRAN}{Evolved Universal Terrestrial Access Network}
\newacronym{faas}{FaaS}{Function-as-a-Service}
\newacronym{fapi}{FAPI}{Functional Application Platform Interface}
\newacronym{fdd}{FDD}{Frequency Division Duplexing}
\newacronym{fdm}{FDM}{Frequency Division Multiplexing}
\newacronym{fdma}{FDMA}{Frequency Division Multiple Access}
\newacronym{fed4fire}{FED4FIRE+}{Federation 4 Future Internet Research and Experimentation Plus}
\newacronym{fir}{FIR}{Finite Impulse Response}
\newacronym{cir}{CIR}{Channel Impulse Response}
\newacronym{fit}{FIT}{Future \acrlong{iot}}
\newacronym{fpga}{FPGA}{Field Programmable Gate Array}
\newacronym{fr2}{FR2}{Frequency Range 2}
\newacronym{fr1}{FR1}{Frequency Range 1}
\newacronym{fs}{FS}{Fast Switching}
\newacronym{fscc}{FSCC}{Flow Sharing Congestion Control}
\newacronym{ftp}{FTP}{File Transfer Protocol}
\newacronym{fw}{FW}{Flow Window}
\newacronym{ntn}{NTN}{Non-Terrestrial Networks}
\newacronym{ge}{GE}{Gaussian Elimination}
\newacronym{fr3}{FR3}{Frequency Range 3}
\newacronym{gnb}{gNB}{Next Generation Node Base}
\newacronym{nextg}{NextG}{Next Generation}
\newacronym{gop}{GOP}{Group of Pictures}
\newacronym{gpr}{GPR}{Gaussian Process Regressor}
\newacronym{gpu}{GPU}{Graphics Processing Unit}
\newacronym{gtp}{GTP}{GPRS Tunneling Protocol}
\newacronym{gtpc}{GTP-C}{GPRS Tunnelling Protocol Control Plane}
\newacronym{sca}{SCA}{Successive Convex Approximation}
\newacronym{gtpu}{GTP-U}{GPRS Tunnelling Protocol User Plane}
\newacronym{gtpv2c}{GTPv2-C}{\gls{gtp} v2 - Control}
\newacronym{gw}{GW}{Gateway}
\newacronym{harq}{HARQ}{Hybrid Automatic Repeat reQuest}
\newacronym{hetnet}{HetNet}{Heterogeneous Network}
\newacronym{hh}{HH}{Hard Handover}
\newacronym{hol}{HOL}{Head-of-Line}
\newacronym{hqf}{HQF}{Highest-quality-first}
\newacronym{hss}{HSS}{Home Subscription Server}
\newacronym{http}{HTTP}{HyperText Transfer Protocol}
\newacronym{ia}{IA}{Initial Access}
\newacronym{iab}{IAB}{Integrated Access and Backhaul}
\newacronym{ic}{IC}{Incident Command}
\newacronym{ietf}{IETF}{Internet Engineering Task Force}
\newacronym{imsi}{IMSI}{International Mobile Subscriber Identity}
\newacronym{imt}{IMT}{International Mobile Telecommunication}
\newacronym{iot}{IoT}{Internet of Things}
\newacronym{ip}{IP}{Internet Protocol}
\newacronym{itu}{ITU}{International Telecommunication Union}
\newacronym{kpi}{KPI}{Key Performance Indicator}
\newacronym{kpm}{KPM}{Key Performance Metric}
\newacronym{kvm}{KVM}{Kernel-based Virtual Machine}
\newacronym{los}{LOS}{Line-of-Sight}
\newacronym{lsm}{LSM}{Link-to-System Mapping}
\newacronym{lstm}{LSTM}{Long Short Term Memory}
\newacronym{lte}{LTE}{Long Term Evolution}
\newacronym{lxc}{LXC}{Linux Container}
\newacronym{m2m}{M2M}{Machine to Machine}
\newacronym{mac}{MAC}{Medium Access Control}
\newacronym{manet}{MANET}{Mobile Ad Hoc Network}
\newacronym{mano}{MANO}{Management and Orchestration}
\newacronym{mc}{MC}{Multi-Connectivity}
\newacronym{mcc}{MCC}{Mobile Cloud Computing}
\newacronym{mchem}{MCHEM}{Massive Channel Emulator}
\newacronym{mcs}{MCS}{Modulation and Coding Scheme}
\newacronym{mec}{MEC}{Multi-access Edge Computing}
\newacronym{mec2}{MEC}{Mobile Edge Cloud}
\newacronym{mfc}{MFC}{Mobile Fog Computing}
\newacronym{mgen}{MGEN}{Multi-Generator}
\newacronym{mi}{MI}{Mutual Information}
\newacronym{mib}{MIB}{Master Information Block}
\newacronym{miesm}{MIESM}{Mutual Information Based Effective SINR}
\newacronym{mimo}{MIMO}{Multiple Input, Multiple Output}
\newacronym{ml}{ML}{Machine Learning}
\newacronym{mlr}{MLR}{Maximum-local-rate}
\newacronym[plural=\gls{mme}s,firstplural=Mobility Management Entities (MMEs)]{mme}{MME}{Mobility Management Entity}
\newacronym{mmtc}{mMTC}{Massive Machine-Type Communications}
\newacronym{mmwave}{mmWave}{millimeter wave}
\newacronym{mpdccp}{MP-DCCP}{Multipath Datagram Congestion Control Protocol}
\newacronym{mptcp}{MPTCP}{Multipath TCP}
\newacronym{mr}{MR}{Maximum Rate}
\newacronym{mrdc}{MR-DC}{Multi \gls{rat} \gls{dc}}
\newacronym{mse}{MSE}{Mean Square Error}
\newacronym{mss}{MSS}{Maximum Segment Size}
\newacronym{mt}{MT}{Mobile Termination}
\newacronym{mtd}{MTD}{Machine-Type Device}
\newacronym{mtu}{MTU}{Maximum Transmission Unit}
\newacronym{mumimo}{MU-MIMO}{Multi-user \gls{mimo}}
\newacronym{mvno}{MVNO}{Mobile Virtual Network Operator}
\newacronym{nalu}{NALU}{Network Abstraction Layer Unit}
\newacronym{nas}{NAS}{Non-Access Stratum}
\newacronym{nbiot}{NB-IoT}{Narrow Band IoT}
\newacronym{nfv}{NFV}{Network Function Virtualization}
\newacronym{nfvi}{NFVI}{Network Function Virtualization Infrastructure}
\newacronym{nic}{NIC}{Network Interface Card}
\newacronym{nlos}{NLOS}{Non-Line-of-Sight}
\newacronym{now}{NOW}{Non Overlapping Window}
\newacronym{nsm}{NSM}{Network Service Mesh}
\newacronym[type=hidden]{nr}{NR}{New Radio}
\newacronym{nrf}{NRF}{Network Repository Function}
\newacronym{nsa}{NSA}{Non Stand Alone}
\newacronym{nse}{NSE}{Network Slicing Engine}
\newacronym{nssf}{NSSF}{Network Slice Selection Function}
\newacronym{o2i}{O2I}{Outdoor to Indoor}
\newacronym{oai}{OAI}{OpenAirInterface}
\newacronym{oaicn}{OAI-CN}{\gls{oai} \acrlong{cn}}
\newacronym{oairan}{OAI-RAN}{\acrlong{oai} \acrlong{ran}}
\newacronym{oam}{OAM}{Operations, Administration and Maintenance}
\newacronym{ofdm}{OFDM}{Orthogonal Frequency Division Multiplexing}
\newacronym{olia}{OLIA}{Opportunistic Linked Increase Algorithm}
\newacronym{omec}{OMEC}{Open Mobile Evolved Core}
\newacronym{onap}{ONAP}{Open Network Automation Platform}
\newacronym{onf}{ONF}{Open Networking Foundation}
\newacronym{onos}{ONOS}{Open Networking Operating System}
\newacronym{oom}{OOM}{\gls{onap} Operations Manager}
\newacronym{opnfv}{OPNFV}{Open Platform for \gls{nfv}}
\newacronym{orbit}{ORBIT}{Open-Access Research Testbed for Next-Generation Wireless Networks}
\newacronym{os}{OS}{Operating System}
\newacronym{oss}{OSS}{Operations Support System}
\newacronym{pa}{PA}{Position-aware}
\newacronym{pase}{PASE}{Prioritization, Arbitration, and Self-adjusting Endpoints}
\newacronym{pawr}{PAWR}{Platforms for Advanced Wireless Research}
\newacronym{pbch}{PBCH}{Physical Broadcast Channel}
\newacronym{pcef}{PCEF}{Policy and Charging Enforcement Function}
\newacronym{pcfich}{PCFICH}{Physical Control Format Indicator Channel}
\newacronym{pcrf}{PCRF}{Policy and Charging Rules Function}
\newacronym{pdcch}{PDCCH}{Physical Downlink Control Channel}
\newacronym{pdcp}{PDCP}{Packet Data Convergence Protocol}
\newacronym{pdsch}{PDSCH}{Physical Downlink Shared Channel}
\newacronym{pdu}{PDU}{Packet Data Unit}
\newacronym{pf}{PF}{Proportionally Fair}
\newacronym{pgw}{PGW}{Packet Gateway}
\newacronym{phich}{PHICH}{Physical Hybrid ARQ Indicator Channel}
\newacronym{phy}{PHY}{Physical}
\newacronym{pmch}{PMCH}{Physical Multicast Channel}
\newacronym{pmi}{PMI}{Precoding Matrix Indicators}
\newacronym{powder}{POWDER}{Platform for Open Wireless Data-driven Experimental Research}
\newacronym{ppo}{PPO}{Proximal Policy Optimization}
\newacronym{ppp}{PPP}{Poisson Point Process}
\newacronym{prach}{PRACH}{Physical Random Access Channel}
\newacronym{prb}{PRB}{Physical Resource Block}
\newacronym{rbg}{RBG}{Resource Block Group}
\newacronym{psnr}{PSNR}{Peak Signal to Noise Ratio}
\newacronym{pss}{PSS}{Primary Synchronization Signal}
\newacronym{pucch}{PUCCH}{Physical Uplink Control Channel}
\newacronym{pusch}{PUSCH}{Physical Uplink Shared Channel}
\newacronym{qam}{QAM}{Quadrature Amplitude Modulation}
\newacronym{qci}{QCI}{\gls{qos} Class Identifier}
\newacronym{qoe}{QoE}{Quality of Experience}
\newacronym{qos}{QoS}{Quality of Service}
\newacronym{quic}{QUIC}{Quick UDP Internet Connections}
\newacronym{rach}{RACH}{Random Access Channel}
\newacronym{ran}{RAN}{Radio Access Network}
\newacronym[firstplural=end to endcess Technologies (RATs)]{rat}{RAT}{end to endcess Technology}
\newacronym{rcn}{RCN}{Research Coordination Network}
\newacronym{rec}{REC}{Radio Edge Cloud}
\newacronym{ra}{RA}{Resource Allocation}
\newacronym{red}{RED}{Random Early Detection}
\newacronym{renew}{RENEW}{Reconfigurable Eco-system for Next-generation End-to-end Wireless}
\newacronym{rf}{RF}{Radio Frequency}
\newacronym{rfc}{RFC}{Request for Comments}
\newacronym{rfr}{RFR}{Random Forest Regressor}
\newacronym{ric}{RIC}{\gls{ran} Intelligent Controller}
\newacronym{rlc}{RLC}{Radio Link Control}
\newacronym{rlf}{RLF}{Radio Link Failure}
\newacronym{rlnc}{RLNC}{Random Linear Network Coding}
\newacronym{rmr}{RMR}{RIC Message Router}
\newacronym{rmse}{RMSE}{Root Mean Squared Error}
\newacronym{rnis}{RNIS}{Radio Network Information Service}
\newacronym{rr}{RR}{Round Robin}
\newacronym{rrc}{RRC}{Radio Resource Control}
\newacronym{rrm}{RRM}{Radio Resource Management}
\newacronym{rru}{RRU}{Remote Radio Unit}
\newacronym{rs}{RS}{Remote Server}
\newacronym{rsrp}{RSRP}{Reference Signal Received Power}
\newacronym{rsrq}{RSRQ}{Reference Signal Received Quality}
\newacronym{rss}{RSS}{Received Signal Strength}
\newacronym{rssi}{RSSI}{Received Signal Strength Indicator}
\newacronym{rtt}{RTT}{Round Trip Time}
\newacronym{ru}{RU}{Radio Unit}
\newacronym{rw}{RW}{Receive Window}
\newacronym{rx}{RX}{Receiver}
\newacronym{s1ap}{S1AP}{S1 Application Protocol}
\newacronym{sa}{SA}{standalone}
\newacronym{sack}{SACK}{Selective Acknowledgment}
\newacronym{sap}{SAP}{Service Access Point}
\newacronym{sc2}{SC2}{Spectrum Collaboration Challenge}
\newacronym{scef}{SCEF}{Service Capability Exposure Function}
\newacronym{sch}{SCH}{Secondary Cell Handover}
\newacronym{scoot}{SCOOT}{Split Cycle Offset Optimization Technique}
\newacronym{sctp}{SCTP}{Stream Control Transmission Protocol}
\newacronym{sdap}{SDAP}{Service Data Adaptation Protocol}
\newacronym{sdk}{SDK}{Software Development Kit}
\newacronym{sdm}{SDM}{Space Division Multiplexing}
\newacronym{sdma}{SDMA}{Spatial Division Multiple Access}
\newacronym{sdn}{SDN}{Software-defined Networking}
\newacronym{sdr}{SDR}{Software-defined Radio}
\newacronym{seba}{SEBA}{SDN-Enabled Broadband Access}
\newacronym{sgsn}{SGSN}{Serving GPRS Support Node}
\newacronym{sgw}{SGW}{Service Gateway}
\newacronym{si}{SI}{Study Item}
\newacronym{sib}{SIB}{Secondary Information Block}
\newacronym{sinr}{SINR}{Signal to Interference plus Noise Ratio}
\newacronym{sip}{SIP}{Session Initiation Protocol}
\newacronym{siso}{SISO}{Single Input, Single Output}
\newacronym{sla}{SLA}{Service Level Agreement}
\newacronym{sm}{SM}{Service Model}
\newacronym{smf}{SMF}{Session Management Function}
\newacronym{smo}{SMO}{Service Management and Orchestration}
\newacronym{sms}{SMS}{Short Message Service}
\newacronym{smsgmsc}{SMS-GMSC}{\gls{sms}-Gateway}
\newacronym{snr}{SNR}{Signal-to-Noise-Ratio}
\newacronym{son}{SON}{Self-Organizing Network}
\newacronym{sptcp}{SPTCP}{Single Path TCP}
\newacronym{srb}{SRB}{Service Radio Bearer}
\newacronym{srn}{SRN}{Standard Radio Node}
\newacronym{srs}{SRS}{Sounding Reference Signal}
\newacronym{ss}{SS}{Synchronization Signal}
\newacronym{sss}{SSS}{Secondary Synchronization Signal}
\newacronym{st}{ST}{Spanning Tree}
\newacronym{svc}{SVC}{Scalable Video Coding}
\newacronym{tb}{TB}{Transport Block}
\newacronym{tcp}{TCP}{Transmission Control Protocol}
\newacronym{tdd}{TDD}{Time Division Duplexing}
\newacronym{tdm}{TDM}{Time Division Multiplexing}
\newacronym{tdma}{TDMA}{Time Division Multiple Access}
\newacronym{tfl}{TfL}{Transport for London}
\newacronym{tfrc}{TFRC}{TCP-Friendly Rate Control}
\newacronym{tft}{TFT}{Traffic Flow Template}
\newacronym{tgen}{TGEN}{Traffic Generator}
\newacronym{tip}{TIP}{Telecom Infra Project}
\newacronym{tm}{TM}{Transparent Mode}
\newacronym{to}{TO}{Telco Operator}
\newacronym{tr}{TR}{Technical Report}
\newacronym{trp}{TRP}{Transmitter Receiver Pair}
\newacronym{ts}{TS}{Technical Specification}
\newacronym{tti}{TTI}{Transmission Time Interval}
\newacronym{ttt}{TTT}{Time-to-Trigger}
\newacronym{tx}{TX}{Transmitter}
\newacronym{uas}{UAS}{Unmanned Aerial System}
\newacronym{uav}{UAV}{Unmanned Aerial Vehicle}
\newacronym{udm}{UDM}{Unified Data Management}
\newacronym{udp}{UDP}{User Datagram Protocol}
\newacronym{udr}{UDR}{Unified Data Repository}
\newacronym{ue}{UE}{User Equipment}
\newacronym{uhd}{UHD}{\gls{usrp} Hardware Driver}
\newacronym{ul}{UL}{Uplink}
\newacronym{ap}{AP}{Access Point}
\newacronym{um}{UM}{Unacknowledged Mode}
\newacronym{uml}{UML}{Unified Modeling Language}
\newacronym{upa}{UPA}{Uniform Planar Array}
\newacronym{ura}{URA}{Uniform Rectangular Array}
\newacronym{upf}{UPF}{User Plane Function}
\newacronym{urllc}{URLLC}{Ultra Reliable and Low Latency Communications}
\newacronym{usa}{U.S.}{United States}
\newacronym{usim}{USIM}{Universal Subscriber Identity Module}
\newacronym{usrp}{USRP}{Universal Software Radio Peripheral}
\newacronym{utc}{UTC}{Urban Traffic Control}
\newacronym{vim}{VIM}{Virtualization Infrastructure Manager}
\newacronym{vm}{VM}{Virtual Machine}
\newacronym{vnf}{VNF}{Virtual Network Function}
\newacronym{volte}{VoLTE}{Voice over \gls{lte}}
\newacronym{voltha}{VOLTHA}{Virtual OLT HArdware Abstraction}
\newacronym{vr}{VR}{Virtual Reality}
\newacronym{vran}{vRAN}{Virtualized \gls{ran}}
\newacronym{vss}{VSS}{Video Streaming Server}
\newacronym{wbf}{WBF}{Wired Bias Function}
\newacronym{wf}{WF}{Waterfilling}
\newacronym{wlan}{WLAN}{Wireless Local Area Network}
\newacronym{osm}{OSM}{Open Source \gls{nfv} Management and Orchestration}
\newacronym{pnf}{PNF}{Physical Network Function}
\newacronym{drl}{DRL}{Deep Reinforcement Learning}
\newacronym{rl}{RL}{Reinforcement Learning}
\newacronym{mtc}{MTC}{Machine-type Communications}
\newacronym{osc}{OSC}{O-RAN Software Community}
\newacronym{rc}{RC}{RAN Control}
\newacronym{dqn}{DQN}{Deep Q-Network}
\newacronym{v2x}{V2X}{Vehicle-to-everything}
\newacronym{gbsm}{GBSM}{Geometry-Based Stochastic Model}
\newacronym{gbs}{GBSM}{Geometry-Based Stochastic}
\newacronym{quadriga}{QuaDRiGa}{QUAsi Deterministic RadIo channel GenerAtor}
\newacronym{relu}{ReLU}{Rectified Linear Unit} 
\newacronym{mpc}{MPC}{Multipath Component}
\newacronym{xpr}{XPR}{Cross-polarization Ratio}
\newacronym{lsp}{LSP}{Large Scale Parameter}
\newacronym{ssp}{SSP}{Small Scale Parameter}
\newacronym{fbs}{FBS}{First Bounce Scatterer}
\newacronym{lbs}{LBS}{Last Bounce Scatterer}
\newacronym{d2d}{D2D}{Device-to-Device}
\newacronym{rsu}{RSU}{Road Side Unit}
\newacronym{toa}{ToA}{Time-of-Arrival}
\newacronym{ris}{RIS}{Reconfigurable Intelligent Surface}
\newacronym{aoa}{AoA}{Angle of Arrival}
\newacronym{aod}{AoD}{Angle of Departure}
\newacronym{pl}{PL}{Path-Loss}
\newacronym{noma}{NOMA}{Non-Orthogonal Multiple Access}
\newacronym{sic}{SIC}{Successive Interference Cancellation}
\newacronym{gps}{GPS}{Global Positioning System}
\newacronym{ids}{IDS}{Independent Diffusive Scatterer-based}
\newacronym{ttd}{TTD}{True-Time-Delay}
\newacronym{inw}{INW}{Impedance Network-based}
\newacronym{sir}{SIR}{Signal-to-Interference Ratio}
\newacronym{minlp}{MINLP}{Mixed Integer Non-Linear Programming}
\newacronym{star}{STAR}{Simultaneous Transmitting And Reflecting}
\newacronym{mbs}{MBS}{Macro Base Station}
\newacronym{wsr}{WSR}{Weighted Sum Rate}
\newacronym{ez}{EZ}{Exclusion Zone}
\newacronym{leo}{LEO}{Low Earth Orbit}
\newacronym{sqp}{SQP}{Sequential Quadratic Programming}
\newacronym{svd}{SVD}{Singular Value Decomposition}
\newacronym{kkt}{KKT}{Karush-Kuhn-Tucker}
\newacronym{inr}{INR}{interference-to-noise ratio}
\newacronym{jfi}{JFI}{Jain's Fairness Index}
\newacronym{pdf}{PDF}{Probability Density Function}
\newacronym{dof}{DOF}{Degrees of Freedom}
\newacronym{mu}{MU}{Multi-User}
\newacronym{epfd}{EPFD}{Equivalent Power Flux-Density}
\newacronym{fss}{FSS}{Fixed-Satellite Service}

\glsdisablehyper

\newcommand{\R}{\Delta\textrm{RSS}}

\usepackage{tikzpagenodes,etoolbox}
\usetikzlibrary{calc}
\usepackage[contents={}]{background}
\AddEverypageHook{%
\ifnumequal{\thepage}{1}{%
 \tikz[remember picture,overlay]{%
     \node[draw,
     text width=0.95\textwidth,
     font=\footnotesize
     ]
     at ($(current page header area) - (0,5pt)$)
     {%
   This paper has been accepted for publication at the IEEE International Symposium on Dynamic Spectrum Access Networks (IEEE DYSPAN), 2026. This is the authors' accepted version of the article. The final version published by IEEE is: M. Tsampazi, P. Testolina, M. Polese, T. Melodia, "Satellite-Terrestrial Spectrum Sharing in FR3 through QoS-Aware Power Control and Spatial Nulling," Proc. of the IEEE International Symposium on Dynamic Spectrum Access Networks (IEEE DYSPAN), Washington, DC, United States, May 2026.
     };
 }%
}{}%end ifnumequal
}

\begin{document}

% \title{Coexistence Strategies for Terrestrial-Satellite Spectrum Sharing in the FR3 Upper Mid-Band\vspace{-.3cm}}
% \title{\title{Power Control vs. Nulling for Satellite-Terrestrial Spectrum Sharing in FR3: A Fairness-Driven Analysis\vspace{-.3cm}}\vspace{-.3cm}}
% \title{Performance Evaluation of Interference Mitigation for Satellite-Terrestrial FR3 Spectrum Sharing\vspace{-.3cm}}
% \title{A Comparative Analysis of Interference Supression Techniques for Satellite-Terrestrial FR3 Coexistence\vspace{-.3cm}}
% \title{Terrestrial-Satellite Spectrum Sharing in FR3:\\ A Fairness Perspective\vspace{-.3cm}}
% \title{Quantifying Fairness-Efficiency Trade-offs in Satellite-Terrestrial FR3 Spectrum Sharing\vspace{-.3cm}}
% \title{Fairness-Constrained Satellite-Terrestrial Spectrum Sharing in FR3: Power Control vs. Spatial Nulling\vspace{-.3cm}}
% \title{Power Control vs. Nulling for Satellite-Terrestrial FR3 Coexistence: A Quantitative Fairness Analysis\vspace{-.3cm}}
% \title{Satellite-Terrestrial Coexistence in FR3: Energy-Efficient Power Control vs. Spatial Nulling\vspace{-.3cm}}
% \title{Satellite-Terrestrial Coexistence in FR3: Comparative Analysis of Power Control and Spatial Nulling\vspace{-.3cm}}

% \title{QoS-Aware Power Control and Spatial Nulling for Satellite-Terrestrial Coexistence in FR3\vspace{-.3cm}}

\title{Satellite-Terrestrial Spectrum Sharing in FR3 through QoS-Aware Power Control and Spatial Nulling\vspace{-.3cm}}

\author{\IEEEauthorblockN{Maria Tsampazi, Paolo Testolina, Michele Polese, Tommaso Melodia}
\IEEEauthorblockA{Institute for Intelligent Networked Systems, Northeastern University, Boston, MA, U.S.A.\\
E-mail: \{tsampazi.m, p.testolina, m.polese, t.melodia\}@northeastern.edu}
\thanks{This work was supported in part by the U.S.\ National Science Foundation under grants CNS-2332721 and CNS-2434081.}
}

\makeatletter
\patchcmd{\@maketitle}
  {\addvspace{0.5\baselineskip}\egroup} %0.5
  {\addvspace{-2.38\baselineskip}\egroup} %-1.5
  {}
  {}
\makeatother

\maketitle
% \glsresetall

\begin{abstract}
\gls{fr3}, encompassing frequencies between $7.125$ and $24.25$~GHz, is an emerging frequency band for \gls{6g} applications. The upper mid-band, as it is frequently referred to, represents the sweet spot between coverage and capacity, providing better range than mmWaves and higher bandwidth than the sub-6~GHz band. Despite these advantages, the spectrum is already occupied by incumbent systems such as satellites (e.g., Starlink), and sharing it with terrestrial cellular applications results in spectrum conflicts, only exacerbating the existing spectrum scarcity. This article investigates the impact of two state-of-the-art methods, namely \gls{qos}-Aware Power Control and Interference Nulling, as well as their joint application, on interference mitigation toward non-terrestrial links while maintaining acceptable \gls{qos} on terrestrial networks.
Our simulation results demonstrate the advantages and disadvantages of each method, pinpointing how interference nulling can maintain high average performance and how power control is more appropriate for risk-averse scenarios to enhance fairness in terrestrial \gls{qos}. Finally, we showcase how the two can complement each other to enhance fairness in terrestrial \gls{qos} and increase the \gls{gnb}'s energy efficiency, while suppressing interference toward incumbents.
\end{abstract}

\glsresetall
\glsunset{nextg}
\glsunset{ris}
\glsunset{ran}
\glsunset{enb}
\glsunset{embb}
\glsunset{urllc}
\glsunset{ric}
\glsunset{usrp}
\glsunset{6g}
\glsunset{fr3}
\glsunset{gnb}
\glsunset{qos}

\begin{IEEEkeywords}
FR3, Spectrum Sharing, 6G, QoS-Aware Power Control, Interference Nulling
\end{IEEEkeywords}

% \vspace{-0.15cm}
\section{Introduction}\label{intro}
\gls{fr3} has been identified as the \textit{golden band} for \gls{6g}, as it satisfies both capacity and coverage requirements~\cite{bazzi2025upper}. However, the rapid deployment of dense satellite communication systems introduces new challenges that necessitate robust interference management strategies. In particular, coexistence between terrestrial and non-terrestrial networks remains a critical issue for the successful commercial deployment of \gls{6g} systems. Therefore, the implementation of advanced interference management techniques is essential.

The challenge of interference mitigation in coexistence scenarios has been extensively studied in the literature~\cite{heydarishahreza2024spectrum}. Proposed strategies range from the use of highly directional antennas to adaptive beamforming and null steering designed to suppress radiation in undesired directions. Furthermore, spectrum masking, power control, and diversity techniques are commonly employed to minimize interference effects~\cite{agarwal2023coexistence}. While these approaches are effective for managing interference in the \gls{fr3}, they often face practical challenges related to hardware limitations and limited spectrum availability~\cite{agarwal2023coexistence}.

Specifically, in~\cite{niloy2024ascent, niloy2023interference}, the authors underscore the importance of site-specific deployment strategies and context-aware policies in ensuring efficient coexistence between terrestrial and satellite systems. In~\cite{niloy2024ascent} a spectrum-sharing scenario is examined in which a terrestrial network operating in the upper mid-band interferes with incumbent satellite communications. Interference mitigation is achieved via a closed-loop feedback mechanism that enables real-time adaptation of spectrum-sharing policies. This approach balances satellite protection with efficient spectrum utilization by dynamically adjusting the \gls{ez} radius whenever interference levels exceed predefined thresholds. In~\cite{niloy2023interference} spectral overlap is discussed with a focus on \gls{los} links and angular alignment between the \gls{gnb} and satellite \glspl{rx}, which are identified as critical factors affecting interference. In~\cite{wu2023space} the authors propose an optimization framework to maximize the \gls{wsr} in multi-\gls{ue} multibeam \gls{leo} satellite networks by optimizing the beamforming weights to balance signal quality and interference mitigation. The work in~\cite{ma2023resource} optimizes beamforming and power in dense \gls{leo} satellite networks to maximize system fairness. Beamforming vectors are calculated using \gls{sca} to maximize the minimum \gls{ue} rate, while satisfying power and interference constraints. If power limits are exceeded, power scaling is applied to adjust the beamforming vectors. In~\cite{kang2024terrestrial}, the authors focus on the trade-off between terrestrial \gls{snr} maximization and interference suppression, represented by the tunable parameter~$\lambda$, and solve it as a joint maximization--minimization problem. Finally, in~\cite{wadaskar2025satellite}, the authors address satellite-terrestrial coexistence in \gls{fr3} by designing hybrid \gls{ttd} array precoders at the \gls{gnb} that steer wideband nulls toward satellites while preserving main-lobe gain for terrestrial \glspl{ue}.

% \vspace{-0.15cm}
\subsection{Contributions and Outline} \label{Section IB}
\begin{figure}[h!]
  \centering
  \includegraphics[width=\columnwidth, height=4.85cm, keepaspectratio]{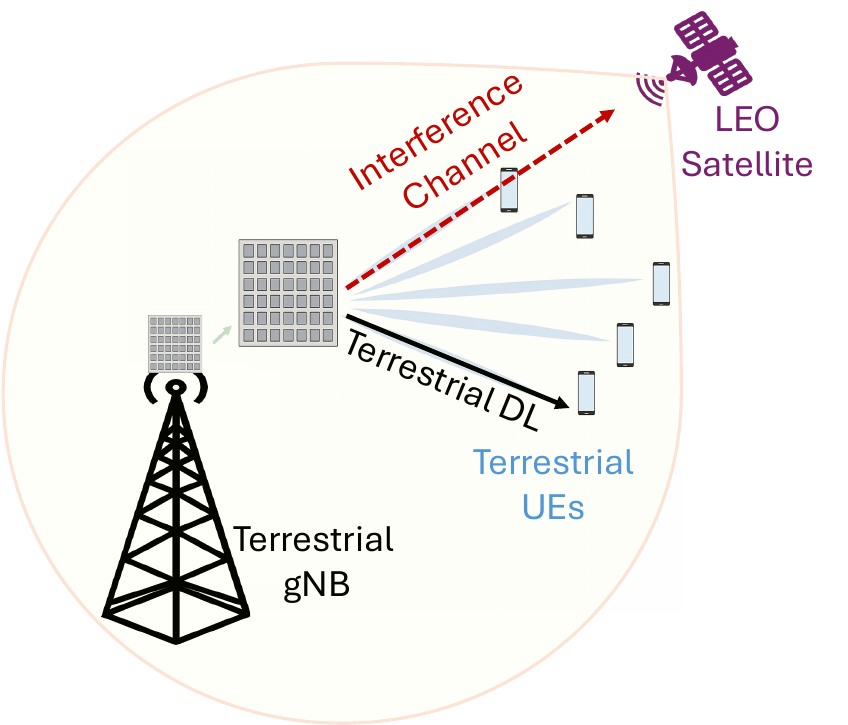}
  \setlength\abovecaptionskip{-.2cm}
  \caption{Interference from terrestrial cellular \gls{dl} transmissions to satellite \gls{ul} in terrestrial-\gls{leo} satellite coexistence scenario.}
  \label{fig:coexistence-top}
  \vspace{-0.55cm}
\end{figure}

However, despite the aforementioned contributions, prior work on spectrum sharing and coexistence has not thoroughly compared widely adopted interference suppression techniques, such as power control and spatial nulling. In this article, we aim to address this gap by considering the worst-case interference caused to satellites, which results from \gls{dl} transmissions in the terrestrial network leaking into the satellite \gls{ul}~\cite{kang2024terrestrial}, as shown in Fig.~\ref{fig:coexistence-top}. Additionally, contrary to prior works~\cite{wadaskar2025satellite, kang2024terrestrial}, which focus on a limited number of satellites (e.g., up to $10$ satellites within a constellation) sharing spectrum with terrestrial cellular networks, we consider deployments involving denser constellations. This allows us to evaluate the impact of interference suppression strategies on scalable coexistence scenarios, thereby pinpointing their strengths and limitations.
Our contributions can be summarized as follows:
\begin{itemize}
    \item We systematically investigate the impact of Interference Nulling on the trade-off between terrestrial \gls{qos} and interference suppression at \gls{leo} satellites.
    \item We formulate an optimization problem to suppress interference at the incumbents by minimizing the \gls{gnb}'s \gls{dl} transmit power, subject to throughput and \gls{inr} constraints.
    \item We provide a comparative analysis between power control strategies and interference nulling techniques, pinpointing the strengths and limitations of each method while identifying suggested use-cases.
    \item We propose a joint interference nulling and power control strategy and benchmark it against standalone baselines.
   \item We conduct our analysis in the context of dense satellite constellations, considering $40$ \gls{leo} satellites to ensure the practical applicability of the proposed solutions under challenging scenarios.
    \item We also evaluate the system's performance using the energy efficiency \gls{kpm}, a critical metric for designing \gls{6g} terrestrial networks.
\end{itemize}

The remainder of this paper is organized as follows. Section~\ref{system model} describes the system model, and Section~\ref{spectrum-strategies} presents the spectrum sharing strategies. Section~\ref{expsetup} details the simulation setup, Section~\ref{kpms} introduces the \glspl{kpm} used for performance evaluation, and Section~\ref{perfevaluation} discusses the simulation results. Finally, Section~\ref{conculsion} concludes the paper and outlines directions for future work.

\section{System Model}\label{system model}
\vspace{-0.5cm}
We focus on a topology comprising a terrestrial \gls{gnb} serving $K$ \glspl{ue} in the \gls{dl}. The \gls{gnb} is equipped with an $N_{\text{az}} \times N_{\text{el}}$ uniform planar array with $N_t = N_{\text{az}} \cdot N_{\text{el}}$ transmit antennas, while each \gls{ue} employs an $N_{\text{az}}^{(R)} \times N_{\text{el}}^{(R)}$ array with $N_r = N_{\text{az}}^{(R)} \cdot N_{\text{el}}^{(R)}$ receive antennas. Both the \gls{gnb} and the \glspl{ue} employ beamforming, selecting transmit and receive beamforming vectors $\boldsymbol{w}_t$ and $\boldsymbol{w}_r$, respectively. It is noted that we employ time-division scheduling, where the \gls{gnb} serves one \gls{ue} per time resource and computes beamforming vectors $\boldsymbol{w}_t$ and $\boldsymbol{w}_r$ for the scheduled \gls{ue}. We additionally consider $N_{\text{sat}}$ \gls{leo} satellites visible from the \gls{gnb} location, sharing the same spectrum. Each satellite is assumed to have a beam focused on the same area covered by the terrestrial \gls{gnb}, making the satellite susceptible to interference from terrestrial transmissions. Finally, the \gls{mimo} channel matrix between the \gls{gnb} and \gls{ue} $k$ is denoted as $\boldsymbol{H}_{\text{ter}}^{k} \in \mathbb{C}^{N_r \times N_t}$, for each \gls{ue} $k \in \{1, \ldots, K\}$, while $\boldsymbol{h}_j \in \mathbb{C}^{N_t \times 1}$ denotes the \gls{mimo} channel vector from the \gls{gnb} to the $j$-th satellite, where $j = 1, \ldots, N_{\text{sat}}$. The time interval during which satellite trajectories are visible to the \gls{gnb} is given as $N_{\text{slots}}$. Therefore, at time slot $t_i$ and for user $k$, the terrestrial channel matrix is denoted as $\mathbf{H}_{\text{ter}}^{k}(t) \in \mathbb{C}^{N_r \times N_t}$.

\section{Terrestrial-Satellite Spectrum Sharing Strategies}\label{spectrum-strategies}

We begin by comparing two state-of-the-art methods, namely Interference Nulling and \gls{qos}-Aware Power Control, to investigate their impact on terrestrial network \gls{qos}, while ensuring that the interference at the satellite remains below the \gls{itu} reference levels in satellite--terrestrial spectrum sharing scenarios. Finally, we propose and evaluate a new coexistence method that combines Interference Nulling and \gls{qos}-Aware Power Control. The performance of the proposed algorithm is systematically evaluated and compared against the standalone methods. 

\subsection{Interference Nulling}\label{interf-null}

The first algorithm we consider is Interference Nulling~\cite{kang2024terrestrial}, where the beamformed gain at the terrestrial \gls{ue} is traded off against the interference leakage toward the satellites via a regularization parameter, $\lambda \geq 0$. The solutions to this optimization problem are the terrestrial transmit and receive beamforming vectors $\boldsymbol{w}_t^{\text{opt}}$ and $\boldsymbol{w}_r^{\text{opt}}$, respectively, as detailed in~\eqref{eq:nulling-opt}:

\begin{equation}\label{eq:nulling-opt}
\begin{aligned}
\boldsymbol{w}_r^{\text{opt}}, \boldsymbol{w}_t^{\text{opt}} &= \arg\max_{\boldsymbol{w}_r, \boldsymbol{w}_t} \left[ |\boldsymbol{w}_r^{\text{H}} \tilde{\boldsymbol{H}}_{\text{ter}} \boldsymbol{w}_t|^2 - \lambda \sum_{j=1}^{N_{\text{sat}}} \|\tilde{\boldsymbol{h}}_j^{\text{H}} \boldsymbol{w}_t\|^2 \right] \\
&\text{subject to\quad} \|\boldsymbol{w}_t\|^2 = 1 \quad \text{and} \quad \|\boldsymbol{w}_r\|^2 = 1.
\end{aligned}
\end{equation}

\noindent
$\tilde{\boldsymbol{H}}_{\text{ter}}$ and $\tilde{\boldsymbol{h}}_j$ denote the normalized versions of the \gls{mimo} multi-path channels $\boldsymbol{H}_{\text{ter}}$ and $\boldsymbol{h}_j$, respectively, as normalizing the channel matrices compensates for the substantial path loss difference between terrestrial and non-terrestrial (i.e., satellite) links, thereby maintaining $\lambda$ within a manageable range. To solve the optimization problem in~\eqref{eq:nulling-opt}, we follow the approach in~\cite{kang2024terrestrial}, where $\boldsymbol{w}_r^{\text{opt}}$ is selected as the maximum left singular vector of $\tilde{\boldsymbol{H}}_{\text{ter}}$, and $\boldsymbol{w}_t^{\text{opt}}$ is chosen as the maximum eigenvector of $\boldsymbol{M} = \tilde{\boldsymbol{H}}_{\text{ter}}^{\text{H}} \boldsymbol{w}_r \boldsymbol{w}_r^{\text{H}} \tilde{\boldsymbol{H}}_{\text{ter}} - \lambda \sum_{j=1}^{N_{\text{sat}}} \tilde{\boldsymbol{h}}_j \tilde{\boldsymbol{h}}_j^{\text{H}}$. It is noted that performing \gls{svd} on $\boldsymbol{H}_{\text{ter}}$ is equivalent to performing \gls{svd} on $\boldsymbol{M}$ for $\lambda = 0$, which is also equivalent to applying eigenvalue decomposition to $\boldsymbol{M}$ for $\lambda = 0$. %in order to obtain the beamforming vector $\boldsymbol{w}_t$.
Therefore, all the aforementioned techniques yield the same solution for the \gls{gnb}'s transmit beamforming vector (i.e., $\boldsymbol{w}_t$) when only beamforming is applied, without accounting for signal leakage toward the satellites. This equivalence serves as a numerical verification of our implementation, confirming that the framework from~\cite{kang2024terrestrial} correctly reduces to standard \gls{svd}-based beamforming when $\lambda = 0$.

\noindent
\subsection{\gls{qos}-Aware Power Control}\label{qos-pc}
In this section, we propose and formulate a power control optimization problem that minimizes the \gls{gnb}'s transmit power while maintaining terrestrial \gls{qos}, in order to ensure satellite protection. We focus on the optimization of the \gls{dl} transmit power of a single \gls{gnb}, and we assume that power allocation decisions are made in a time-division manner, where, at each time slot, a single \gls{ue} is scheduled. The interference-plus-noise levels experienced in the terrestrial network due to nearby operating \glspl{gnb} are given in~\eqref{eq:terr-snir}:
\begin{equation}\label{eq:terr-snir}
{I_{i}^{\text{ter}}=\sum_{n \neq i} G_n P_n + \sigma^2},
\end{equation}

\noindent
where $\sigma^2$ is the \gls{awgn} power and $\sum_{n \neq i} G_n P_n$ is the sum of inter-cell interference power from all neighboring cell sites. The \gls{sinr} for each \gls{gnb}-\gls{ue} link is given in~\eqref{eq:sinr-ter}:

\begin{equation}\label{eq:sinr-ter}
\text{\gls{sinr}}_i^{\text{ter}} = \frac{P_i \left| \boldsymbol{w}_r\boldsymbol{H}_{\text{ter}} \boldsymbol{w}_t \right|^2}{I_{i}^{\text{ter}}},
\end{equation}

\noindent
while the achievable rate for bandwidth $W$ is given by the Shannon Capacity formula as follows in~\eqref{eq:rate}:

\begin{equation}\label{eq:rate}
R_i(P_i) = W \log_2\left(1 + \text{SINR}_i^{\text{ter}}\right).
\end{equation}

\textbf{\textit{\gls{qos}-Aware Utility Function.}} The utility function we consider in our optimization problem is given in~\eqref{eq:PC_obj} as:
\begin{equation}\label{eq:PC_obj}
U_i(P_i) = \frac{W \cdot \left(1 - e^{-\alpha \, \text{SINR}_i^{\text{ter}}}\right)^{M}}{P_i},
\end{equation}

\noindent
and is defined based on a network economics approach~\cite{tsiropoulou2015combined,tsampazi2025system}, capturing the trade-off between the terrestrial network's \gls{qos} satisfaction and the transmission power invested by the \gls{gnb}. This formulation ensures that the \gls{gnb} selects transmission power levels that strike a balance between meeting \gls{qos} demands in terms of achieved terrestrial \gls{sinr} and power savings. In detail, due to the sigmoidal nature of the utility function numerator, small increases in transmission power initially improve the utility by enhancing the \gls{sinr}. However, as the \gls{sinr} increases, the utility function begins to saturate, meaning that further increases in transmission power yield diminishing returns in \gls{qos} satisfaction. 
Beyond this saturation point, increasing power levels further provides little improvement in utility, effectively preventing unnecessary power expenditure. Consequently, the sigmoidal function drives power allocation toward an optimum where the desired \gls{qos} is achieved with less power consumption. Finally, $a$ and $M$ are control parameters governing the slope of the utility function. To ensure that the utility function remains quasi-concave with an interior optimum over the feasible power range, these parameters are tuned based on the interference-plus-noise levels experienced in the terrestrial network~\cite{tsiropoulou2015combined}, as given in~\eqref{eq:terr-snir}.

% \noindent
 For satellite $j$ and a \gls{gnb} employing transmit beamforming vector $\boldsymbol{w}_t$, the \gls{inr} is calculated as follows in~\eqref{eq:inr}:
\begin{equation}\label{eq:inr}
\text{INR}_j = P_{i} + 10 \log_{10}|\boldsymbol{w}_t^H \boldsymbol{h}_{j}|^2 + G/T - L_a - 10 \log_{10}(W\cdot k_B),
\end{equation}
\noindent
where %$P_i$ is the \gls{gnb}'s transmit power, $\mathbf{h}_{j}$ is the multi-path interference channel between the \gls{gnb} and satellite $j$, 
$G/T$ is the satellite antenna gain-to-noise-temperature ratio, $L_a$ accounts for atmospheric and scintillation losses~\cite{3GPP38821}, and $k_B$ is the Boltzmann constant.

% \noindent
Finally, the optimization problem is ultimately formulated in~\eqref{eq:power_control_obj}--\eqref{eq:power_control_power}: 
\begin{subequations}
\label{eq:power_control_optimization}
\begin{align}
\underset{P_i}{\text{maximize}} \quad & U_i(P_i)\label{eq:power_control_obj}  \\
% \max_{P_i} \, & U_i(P_i) = \frac{W \cdot\left(1-e^{-\alpha \text{SINR}_i^{\text{ter}}}\right)^{M}}{P_i} \label{eq:power_control_obj} \\
\text{subject to} \quad & R_i(P_i) \geq \epsilon R_i^{\text{max}}, \label{eq:power_control_rate} \\
& \text{INR}_j(P_i) \leq \text{INR}_{\max}, \label{eq:power_control_inr} \\
& P_i^{\min} \leq P_i \leq P_i^{\max}, \quad \forall i \in I, \label{eq:power_control_power}
\end{align}
\end{subequations}
% \begin{subequations}
% \label{eq:power_control_optimization}
% \begin{align}
% % \underset{P_i}{\text{maximize}} \, & U_i(P_i)\label{eq:power_control_obj}  \\
% \max_{P_i} \, & U_i(P_i) = \frac{W \cdot\left(1-e^{-\alpha \text{SINR}_i^{\text{ter}}}\right)^{M}}{P_i} \label{eq:power_control_obj} \\
% \phantom{\max_{P_i} \,} & \text{subject to} \quad R_i(P_i) \geq \epsilon R_i^{\text{max}}, \label{eq:power_control_rate} \\
% \phantom{\max_{P_i} \,} & \phantom{\text{s.t.}} \quad \text{INR}_j(P_i) \leq \text{INR}_{\max}, \label{eq:power_control_inr} \\
% \phantom{\max_{P_i} \,} & \phantom{\text{s.t.}} P_i^{\min} \leq P_i \leq P_i^{\max}, \quad \forall i \in I, \label{eq:power_control_power}
% \end{align}
% \end{subequations}

\noindent
and pertains to the selection of the power budget at the \gls{gnb} side that satisfies the operator's throughput constraints, while providing maximum interference reduction to the satellites, according to \gls{itu}'s requirements. Therefore, for each link $i \in I$, the \gls{gnb}-\gls{ue} link is allocated the full transmit power budget of the \gls{gnb}, denoted as $P_i$, and which satisfies the constraints defined in~\eqref{eq:power_control_rate}-\eqref{eq:power_control_power}. $U_i(P_i)$ is the \gls{gnb}'s utility function and $R_i(P_i)$ is the achievable rate for power level $P_i$.
% and is given by the Shannon Capacity formula denoted as $R_i = W \cdot \log_2\left(1 + \text{SINR}_i^{\text{ter}}\right)$. 
\noindent
Finally, $P_{\min}$ and $P_{\max}$ are the minimum and maximum transmit power level of the \gls{gnb} respectively, $\text{INR}_{\max}$ is the maximum interference level leaked toward the satellite as permitted by the \gls{itu} (e.g., $-6$~dB~\cite{ITU-R-RS.2017-0}), $\epsilon \in (0,1]$ is the throughput threshold ratio (e.g., $\epsilon = 0.9$ for maintaining $90\%$ of the maximum theoretical throughput $R_i(P_i)$ attained at $P_{i}=P_{\max}$).

 % To restore the bell shape and ensure an interior optimum, the parameter $a$ must be scaled proportionally with $N_i$ (i.e., $a \propto N_i$), such that the sigmoid transitions smoothly across the power range—specifically, from approximately $f(\text{SINR}_i) \approx 0.1$ at $P_{\min}$ to $f(\text{SINR}_i) \approx 0.99$ at $P_{\max}$. This tuning ensures the necessary trade-off between the increasing sigmoidal numerator and the decreasing power penalty term $1/P_i$, yielding a well-defined interior optimum that balances QoS satisfaction and power expenditure.

\subsection{Joint Interference Nulling and QoS-Aware Power Control}\label{joint-opt}

\begin{figure}[h!]
  \centering
  \includegraphics[width=\columnwidth, height=3.4cm, keepaspectratio]{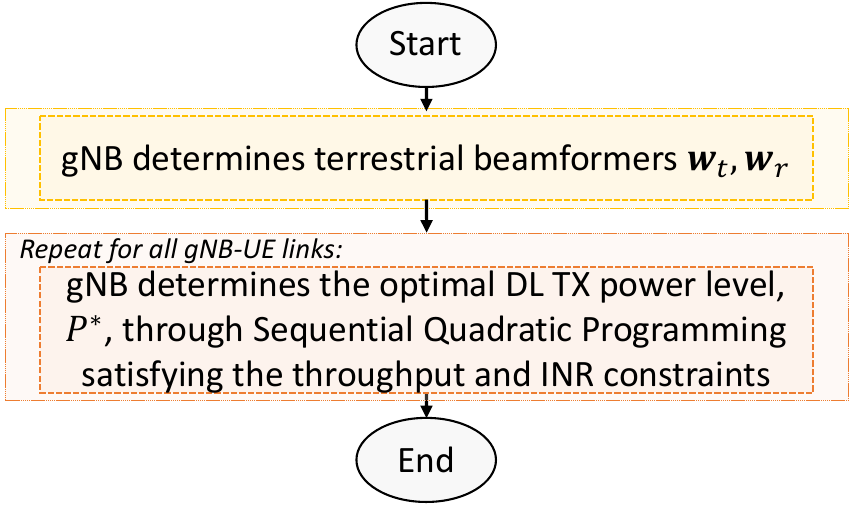}
  \setlength\abovecaptionskip{-.2cm}
  \caption{End-to-end resource management framework overview.}
  \label{fig:flowchart}
  \vspace{-0.6cm}
\end{figure}

The algorithm formulated in~\eqref{eq:nulling-opt} captures the trade-off between ensuring \gls{qos} for the terrestrial network and minimizing interference leakage toward the satellites through the regularization parameter $\lambda$, while the formulation in~\eqref{eq:power_control_obj}-\eqref{eq:power_control_power} ensures that only power budgets satisfying the throughput constraint are considered, and among those, the one providing maximum interference reduction to the satellite is ultimately selected. The formulation for the joint selection of the power level at the \gls{gnb} and the beamforming vectors is given in Algorithm~\ref{alg:joint-beamforming-power-control} and summarized in Fig.~\ref{fig:flowchart}.

\begin{algorithm}[h!!]
\caption{Joint Interference Nulling and \gls{qos}-Aware Power Control}
\label{alg:joint-beamforming-power-control}
\begin{algorithmic}[1]
\REQUIRE Channel matrix $\boldsymbol{H}_{\text{ter}}$ for scheduled \gls{ue} $k$, satellite channels $\{\boldsymbol{h}_j\}_{j=1}^{N_{\text{sat}}}$, regularization parameter $\lambda \geq 0$, utility parameters $\alpha$, $M$, throughput threshold $\epsilon$, INR cap $\text{INR}_{\max}$, power bounds $[P_{\min}, P_{\max}]$, interference-plus-noise $I_i^{\text{ter}}$, bandwidth $W$
\ENSURE Beamforming vectors $\boldsymbol{w}_t^{\text{opt}}$, $\boldsymbol{w}_r^{\text{opt}}$, and power $P_i^{\text{opt}}$
\STATE \textbf{Phase 1: Beamforming Vector Selection}
\STATE Normalize: $\tilde{\boldsymbol{H}}_{\text{ter}} \leftarrow \boldsymbol{H}_{\text{ter}} / \|\boldsymbol{H}_{\text{ter}}\|_F$, \\ $\tilde{\boldsymbol{h}}_j \leftarrow \boldsymbol{h}_j / \|\boldsymbol{h}_j\|_2$, $\forall j \in J$
\STATE Compute \gls{svd}: $\tilde{\boldsymbol{H}}_{\text{ter}} = \boldsymbol{U} \boldsymbol{\Sigma} \boldsymbol{V}^H$
\STATE Assign the optimal receive beamformer as the dominant left singular vector: $\boldsymbol{w}_r^{\text{opt}} \leftarrow \boldsymbol{u}_1$
\STATE Form $\boldsymbol{M} \leftarrow \tilde{\boldsymbol{H}}_{\text{ter}}^H \boldsymbol{w}_r^{\text{opt}} (\boldsymbol{w}_r^{\text{opt}})^H \tilde{\boldsymbol{H}}_{\text{ter}} - \lambda \sum_{j=1}^{N_{\text{sat}}} \tilde{\boldsymbol{h}}_j \tilde{\boldsymbol{h}}_j^H$
\STATE Compute eigendecomposition: $\boldsymbol{M} = \boldsymbol{Q} \boldsymbol{\Lambda} \boldsymbol{Q}^H$
\STATE Set the transmit beamformer to the principal eigenvector: $\boldsymbol{w}_t^{\text{opt}} \leftarrow \boldsymbol{q}_1$
\STATE \textbf{Phase 2: Power Control Optimization}
\STATE Compute the terrestrial channel gain: \\ $G_i^{\text{ter}} \leftarrow |(\boldsymbol{w}_r^{\text{opt}})^H \boldsymbol{H}_{\text{ter}} \boldsymbol{w}_t^{\text{opt}}|^2$
\STATE Compute $R_i^{\text{max}} \leftarrow W \log_2(1 + P_{\max} G_i^{\text{ter}} / I_i^{\text{ter}})$, \\ $R_i^{\min} \leftarrow \epsilon R_i^{\text{max}}$
\STATE Compute the satellite channel gain:\\ $G_j^{\text{sat}} \leftarrow |(\boldsymbol{w}_t^{\text{opt}})^H \boldsymbol{h}_j|^2$, $\forall j \in J$
\STATE Define the Utility function: \\ $f(P_i) \leftarrow -W (1-e^{-\alpha P_i G_i^{\text{ter}} / I_i^{\text{ter}}})^{M} / P_i$
\STATE  Set Throughput constraint: \\ $c_1(P_i) \leftarrow R_i^{\min} - W \log_2(1 + P_i G_i^{\text{ter}} / I_i^{\text{ter}})$
\STATE Set INR constraint: \\ $c_2(P_i) \leftarrow (P_i G_j^{\text{sat}} / N_{\text{sat}}) - 10^{\text{INR}_{\max}/10}$, $\forall j \in J$
\STATE Solve $\min_{P_i} f(P_i)$ subject to $c_1(P_i) \leq 0$, $c_2(P_i) \leq 0$ and  $P_{\min} \leq P_i \leq P_{\max}$ via \gls{sqp}
\RETURN $\boldsymbol{w}_t^{\text{opt}}$, $\boldsymbol{w}_r^{\text{opt}}$, $P_i^{\text{opt}}$
\end{algorithmic}
\end{algorithm}

\section{simulation Setup}\label{expsetup}

\subsection{Terrestrial Channel Model}\label{ter-channelmod}

\begin{figure}[h!]
\centering
% \vspace{-0.55cm}
\setlength\abovecaptionskip{-.1cm}
\subfigure[Terrestrial \gls{los} Path Gain as given by QuaDRiGa's \texttt{power\_map} with the 3GPP\_38.901\_RMa\_LOS channel.]{\includegraphics[width=0.48\columnwidth]{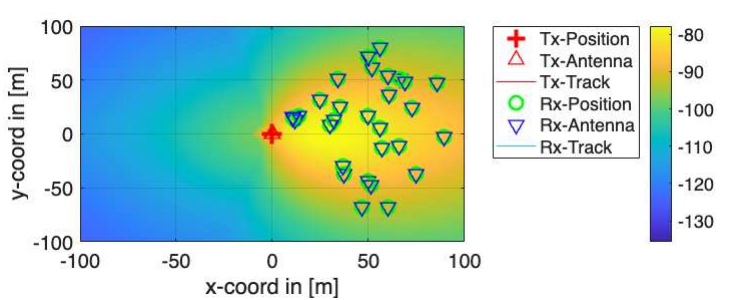}
\label{fig:ter-topology}}
\hfill
\subfigure[Complete topology overview of the terrestrial and \gls{ntn} network from \gls{gnb}'s perspective in \quadriga.]
{\includegraphics[width=0.48\columnwidth]{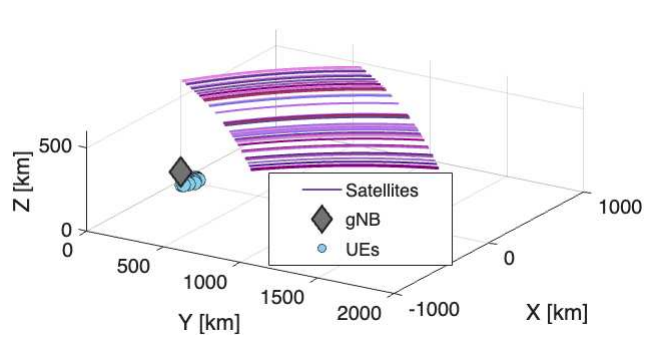}
\label{fig:full-topology}}
\caption{Simulation topology for terrestrial-satellite spectrum sharing.}
\label{fig:topology-combined}
\vspace{-0.5cm}
\end{figure}

Without loss of generality, we assume that all \glspl{ue} are within the coverage of a single \gls{gnb} sector, as shown in Fig.~\ref{fig:ter-topology}, and that all satellites are \textit{visible} from the \gls{gnb}'s sector for most of their trajectory segments.
% We focus on the performance of a single \gls{gnb} as shown in Fig.~\ref{fig:topology-combined}. 
In order to create the terrestrial links, we leverage the \gls{quadriga}~\cite{jaeckel2014quadriga}, a quasi-deterministic geometry-based stochastic channel model simulator compliant with 3GPP TR~38.901 specifications~\cite{3gpp38901}. A detailed tutorial on channel modeling with \quadriga can be found in~\cite{jaeckel2019f,tsampazi2025system}, but full details are omitted in this paper. For the creation of the terrestrial links, we use the parameters defined in Table~\ref{tab:sim_params}. It is noted that the geographical area of the Earth station considered in our simulation is located in central Spain, with geographical coordinates given as $38^\circ51'00.0''\text{N}, 5^\circ00'00.0''\text{W}$.

\subsection{\gls{ntn} Channel Model}\label{ntn-channelmod}
\begin{figure}[h!]
\centering
% \vspace{-0.25cm}
\setlength\abovecaptionskip{-.07cm}
\subfigure[Horsham Township, PA ($C_{in}=95.4$, $D_{in}=-70.57$, $E_{in}=0$, $F_{in}=40.42$)]{\includegraphics[width=0.23\textwidth]{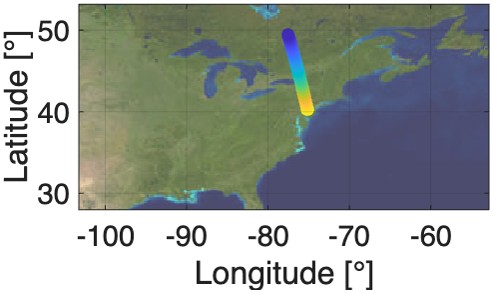}
\label{fig:penns}}
\hfill
\subfigure[Central Spain, EU ($C_{in}=63.4$, $D_{in}=-28.8$, $E_{in}=44.55$, $F_{in}=0$)]{\includegraphics[width=0.23\textwidth]{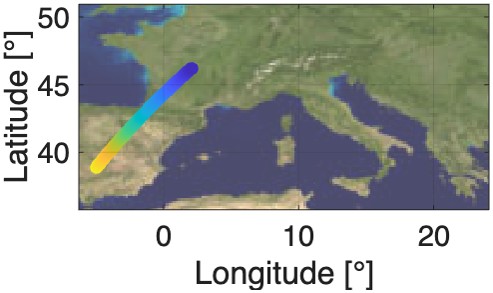}
\label{fig:spain}}
\caption{Satellite overview with \quadriga's \texttt{visualize\_lotlan} in various geographic regions. The trajectories shown correspond to different parameterizations of the orbital elements defined in Listing~\ref{lst:satellite}.}
\label{fig:locations}
\vspace{-0.55cm}
\end{figure}

% \vspace{-0.5em}
For the creation of the non-terrestrial links, we leverage \quadriga's satellite channel modeling approach as it has been discussed in~\cite{jaeckel20225g}. While a descriptive tutorial on satellite channel modeling is omitted, we provide an example of the configuration used in our work to define the satellite's trajectories as follows in Listing~\ref{lst:satellite}, while the configuration of the \gls{ntn} network is provided in Table~\ref{tab:sim_params}. In Fig.~\ref{fig:locations}, we show example satellite orbits created using 
\quadriga's \texttt{qd\_satellite} constructor, initialized with different 
Keplerian orbital elements as defined in Listing~\ref{lst:satellite}. In our work, we consider the satellite constellation corresponding to Fig.~\ref{fig:spain}. For each of the $N_{\text{sat}}$ satellites, and in order to ensure distinct trajectories, we vary their inclination (i.e., Cin of Listing~\ref{lst:satellite}) by an offset parameter~$\Delta$, as  depicted in Fig.~\ref{fig:full-topology}. Finally, the satellite trajectory, given the initialized parameters, is determined using \quadriga's \texttt{orbit\_predictor}.

\begin{lstlisting}[float=h!, floatplacement=b, language=Matlab, 
basicstyle=\ttfamily\scriptsize,
linewidth=\columnwidth,
numbers=left,
numbersep=5pt,
xleftmargin=7pt,
caption={LEO satellite constellation definition in QuaDRiGa.}, 
label={lst:satellite}]
h_qd_sat = qd_satellite(constellation, Ain, Bin, Cin, Din, Ein, Fin);
% Ain: Semimajor axis [km]
% Bin: Orbital eccentricity [0-1]
% Cin: Orbital inclination [deg]
% Din: Ascending node longitude [deg]
% Ein: Argument of periapsis [deg]
% Fin: True anomaly [deg]
\end{lstlisting}

% \vspace{-0.5cm}
\subsection{Considered Constellation}\label{constellation-discussion}
\begin{figure}[h!]
    \centering
    \setlength\abovecaptionskip{-.22cm}
    \resizebox{\columnwidth}{!}{% This file was created by matlab2tikz.
%
\definecolor{mycolor1}{rgb}{0.20000,0.40000,0.80000}%
\definecolor{mycolor2}{rgb}{0.12941,0.12941,0.12941}%
\begin{tikzpicture}

\begin{axis}[%
width=2.5in,
height=0.31in,
at={(1.083in,0.785in)},
scale only axis,
bar shift auto,
clip=false,
xmin=9.95200000000002,
xmax=90.848,
xlabel style={font=\scriptsize},
xlabel={Elevation angle ($\theta$)  [$^\circ$]},
ymin=0,
ymax=0.05,
ylabel style={font=\scriptsize},
ylabel={PDF},
axis background/.style={fill=white},
xmajorgrids,
ymajorgrids,
tick label style={font=\scriptsize},
legend style={legend cell align=left, align=left, font=\scriptsize}
]
\addplot[ybar, bar width=2.56, fill=mycolor1, fill opacity=0.7, draw=black, area legend] table[row sep=crcr] {%
13.28	0.00364583333333333\\
15.84	0.0182291666666667\\
18.4	0.0299479166666667\\
20.96	0.0438802083333333\\
23.52	0.0428385416666667\\
26.08	0.0383463541666667\\
28.64	0.0338541666666666\\
31.2	0.0311848958333333\\
33.76	0.024609375\\
36.32	0.0171875\\
38.88	0.0188802083333334\\
41.44	0.00944010416666666\\
44	0.00820312499999999\\
46.56	0.00904947916666666\\
49.12	0.00683593750000001\\
51.68	0.00722656249999999\\
54.24	0.00781249999999999\\
56.8	0.00546875000000001\\
59.36	0.00624999999999999\\
61.92	0.00520833333333333\\
64.48	0.00436197916666669\\
67.04	0.0033203125\\
69.6	0.003125\\
72.16	0.00279947916666666\\
74.72	0.00286458333333333\\
77.28	0.00227864583333333\\
79.84	0.001171875\\
82.4	0.00149739583333333\\
84.96	0.000520833333333336\\
87.52	0.0005859375\\
};
\addplot[forget plot, color=mycolor2] table[row sep=crcr] {%
9.95200000000002	0\\
90.848	0\\
};
% \addlegendentry{Histogram}

\node[fill=white, below left, align=left, inner sep=1.5, font=\scriptsize, draw=black]
at (rel axis cs:1.0,0.98) {
Mean: $33.0^\circ$, Median: $28.8^\circ$\\
Std: $14.8^\circ$, Min: $13.1^\circ$, Max: $88.6^\circ$
};
\end{axis}
\end{tikzpicture}%
}
    \caption{\gls{pdf} of the satellite elevation angles for the constellation of $N_{\text{sat}} = 40$ satellites considered in this work.}
    \label{fig:elevation-angle-pdf}
    \vspace{-0.3cm}
\end{figure}
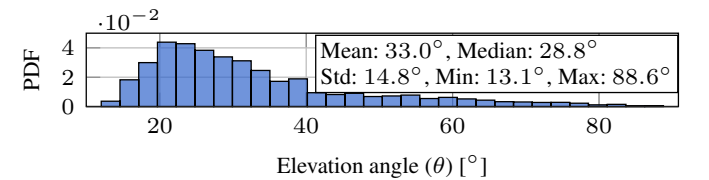
To study the interference leakage from the \gls{gnb} toward the satellite \gls{ul}, we focus on lower elevation angles, as these correspond to scenarios with considerable terrestrial-to-satellite interference. Previous studies~\cite{li2002analytical,kang2024terrestrial} have shown that such angles may lead to increased interference due to the higher antenna gain of the \gls{gnb} in the direction of the satellite \gls{ul}. In addition, the probability of encountering satellites at lower elevation angles is higher, which represents a realistic use-case scenario.
The \gls{pdf} of the satellite elevation angles observed from the considered constellation is shown in Fig.~\ref{fig:elevation-angle-pdf}.

\begin{table}[h]
\caption{Simulation Parameters}
\setlength\abovecaptionskip{-.5cm}
\label{tab:sim_params}
\scriptsize  % Options: \footnotesize, \scriptsize, \tiny
\setlength{\tabcolsep}{4pt}
\begin{tabular}{|c|c|}
\hline
\multicolumn{2}{|c|}{\textbf{Terrestrial Network}} \\
\hline
Carrier frequency  & $7.125$ GHz   \\
\gls{ue} noise figure & $7$~dB \\ 
Cell radius \& Number of \glspl{ue} & $100$~m \& $30$ \\
Antenna array size & $8 \times 8$ (\gls{gnb}) \& $2 \times 1$ (\gls{ue}) \\
Antenna Configuration & \texttt{3gpp-3d}, $12^\circ$ downtilt (\gls{gnb})\\
Antenna height & $50$~m (\gls{gnb}) \& $1.6$~m (\gls{ue}) \\
Channel model & 3GPP 38.901 Rural Macro LOS \\
\hline
\multicolumn{2}{c}{} \\
\hline
\multicolumn{2}{|c|}{\textbf{Satellite Network}} \\
\hline
Number of satellites $(N_{\text{sat}})$  & $40$  \\
 $G/T$ & $13$~dB/K\\
Satellite Antenna Gain \& Altitude & $32$~dBi \& $600$~km\\
Satellite Antenna Configuration & \texttt{parabolic}, $0.25$~m aperture radius\\
Trajectory duration $(N_\text{slots})$ & $150$ (satellite visible to \gls{gnb} throughout)\\
Channel model & QuaDRiGa NTN Rural LOS \\
\hline
\multicolumn{2}{c}{} \\
\hline
\multicolumn{2}{|c|}{\textbf{Power Control Parameters}} \\
\hline
Bandwidth (W) & $30$ MHz \\
 $\text{INR}_\text{max}$ (target \gls{inr}) & $-6$~dB \\
Inter-cell Interference $(I_{i}^{ter})$ & $-73$~dBm \\
$\alpha$ \& $M$ (sigmoid utility parameters) & $10^{-3}$ \& $3$ \\
\gls{gnb} min \& max TX power  & $10$ \& $33$ dBm \\
Throughput thresholds ($\epsilon$) & $\epsilon \in \{0.85, 0.98\}$ \\
\hline
\end{tabular}
\end{table}

\section{Key Performance Evaluation Metrics}\label{kpms}
We introduce the \glspl{kpm} used 
to evaluate the performance of the algorithms. Recall that the goal of this work is to suppress interference toward the satellites, while maintaining acceptable \gls{qos} in the terrestrial network.

We evaluate the \gls{inr} at the satellite, as defined in~\eqref{eq:inr}, to measure the effectiveness of the interference mitigation methods.
%Notably, reducing the \gls{inr} at the satellites to protect the satellite link comes at the cost of terrestrial \gls{qos} due to the beamforming constraints. 
% The \gls{inr} can be directly mapped to the satellite \gls{snr} degradation $\rho^S$ as follows~\cite{ITU_S1432} in Eq.~\eqref{eq:snr_degradation}:
% \begin{equation}
%     \rho^S = 10 \log_{10} \left( 1 + 10^{0.1 \cdot \text{INR}} \right).
%     \label{eq:snr_degradation}
% \end{equation}
For the terrestrial network, we consider the following \glspl{kpm}.

\noindent\textbf{\textit{Terrestrial \gls{rss} degradation.}}
We define the terrestrial \gls{rss} loss at \gls{ue} $k$ as:
\begin{equation}
\label{eq:rss_degradation}
%\rho^{\mathrm{T}}
\Delta\textrm{RSS}_k
= 10\log_{10}\left(\frac{P_{\text{max}} \cdot |\boldsymbol{w}_k^{\mathrm{opt}H}\boldsymbol{H}_{\mathrm{ter}}\boldsymbol{w}_{t,\lambda=0}|^2}{\hat{P}_i \cdot |\boldsymbol{w}_k^{\mathrm{opt}H}\boldsymbol{H}_{\mathrm{ter}}\widehat{\boldsymbol{w}}_{t,\lambda}|^2}\right),
\end{equation}
\noindent
where $\boldsymbol{w}_k^{\mathrm{opt}}$ and $\widehat{\boldsymbol{w}}_{t,\lambda}$ are the \gls{ue} $k$'s and the \gls{gnb} beamforming vectors, respectively, obtained as outlined in Section~\ref{interf-null} with the corresponding $\lambda$ value.
$P_{\text{max}}$ and $\hat{P}_i<P_{\max}$ denote the maximum available and the selected transmit power of the \gls{gnb}, respectively.
The numerator of~\eqref{eq:rss_degradation} corresponds to the best performance case for the terrestrial network, i.e., when no interference mitigation measure is adopted, while the denominator quantifies the impact of the proposed methods (interference nulling and power control) on the terrestrial network \gls{qos}.
Specifically,
\begin{itemize}
\item when evaluating the interference nulling algorithm, without power control, we set $\hat{P}_i=P_{\text{max}}$
\item when evaluating the power control optimization, we consider $\lambda=0$, so $\hat{w}_{t,\lambda}=w_{t,\lambda=0}$ and~\eqref{eq:rss_degradation} reduces to $\Delta\textrm{RSS}_k
= 10\log_{10}\left(P_{\text{max}}/\hat{P}_i\right)$
\item when power control and interference nulling are jointly applied, $\hat{\boldsymbol{w}}_{t,\lambda}$ (for $\lambda>0$) is determined by interference nulling and $\hat{P}_i$ by the power control algorithm
\end{itemize}
% It is further noted that the channel quality of the terrestrial \gls{ue} also varies for different values of $\lambda$, as shown in the results discussed in Section~\ref{perfevaluation}, with larger $\lambda$ parameters resulting in weaker channels. 
$\tilde{\R_k}$ represents the mean \gls{rss} degradation for \gls{ue} $k$, averaged across all $N_\text{slots}$ time slots of the satellites' trajectories.

\textbf{\textit{\gls{jfi}.}} To quantify the fairness of \gls{rss} degradation across the terrestrial \glspl{ue}, we employ the \acrfull{jfi}~\cite{jain1984quantitative}.
For a set of $K$ \glspl{ue} with average \gls{rss} degradation $\{\tilde{\R_1}, \tilde{\R_2}, \ldots, \tilde{\R_K}\}$, the index is defined as:
\begin{equation}
J = \frac{\left(\sum_{k=1}^{K} \tilde{\R_k}\right)^2}{K \cdot \sum_{k=1}^{K} \tilde{\R_k}^2}.
\label{eq:jain_index}
\end{equation}
\noindent
The index ranges from $1/K$ (worst-case fairness, where degradation is concentrated on a single \gls{ue}) to $1$ (perfect fairness, where all \glspl{ue} experience identical degradation).
A value closer to unity indicates a more equitable distribution of \gls{rss} degradation.
This allows us to compare the fairness characteristics of different interference management methods, including interference nulling with various regularization parameters $\lambda$ and power control strategies.

\section{Performance Evaluation}\label{perfevaluation}
In this section, we proceed by evaluating and comparing the optimization methods presented in Section~\ref{spectrum-strategies}, identifying their key advantages and limitations, and investigating how they can complement each other.
It is noted that our simulation findings have been averaged over multiple repetitions. We recall that the parameters considered in this paper for the system-level experimental analysis are given in Table~\ref{tab:sim_params}, while the optimization considers the scheduling decisions of a single \gls{gnb}, given the interference from nearby cells. Multi-cell scheduling interactions are not considered within the scope of this work. Finally, the constrained optimization problem defined in~\eqref{eq:power_control_obj}--\eqref{eq:power_control_power} was solved by leveraging MATLAB's~\cite{MATLAB:2025} \texttt{fmincon} with \gls{sqp}.

\subsection{Impact of Interference Nulling on Beamforming Gain}\label{eval1}

\begin{figure*}[t!]
\centering
%
%\hspace{-0.024\textwidth}
\subfigure[]{\includegraphics[height=3.5cm, width=4.4cm]{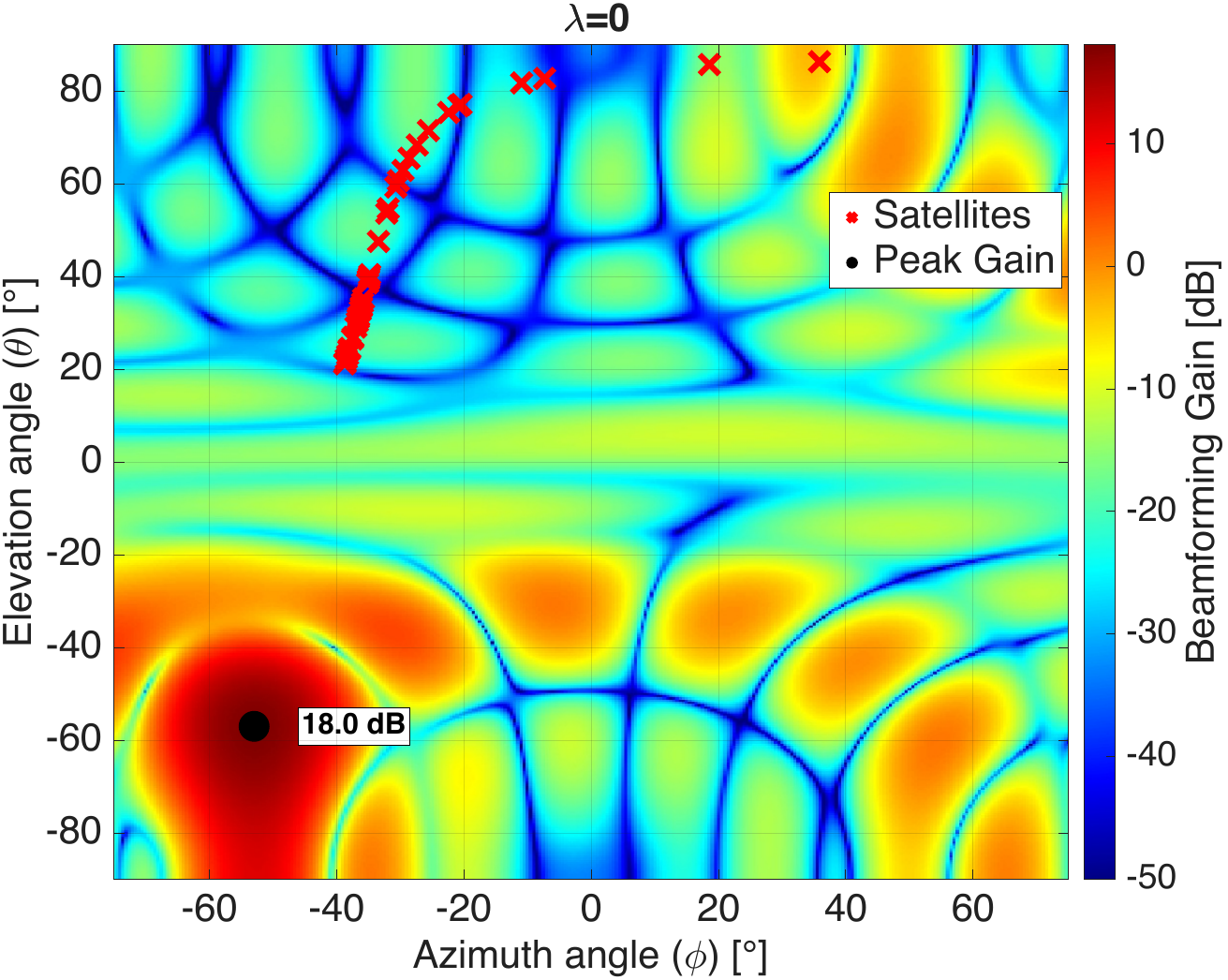}
\label{fig:bf-sim-start-0}}
%
%\hfil
%
\hspace{-0.25cm}
\subfigure[]{\includegraphics[height=3.5cm, width=4.4cm]{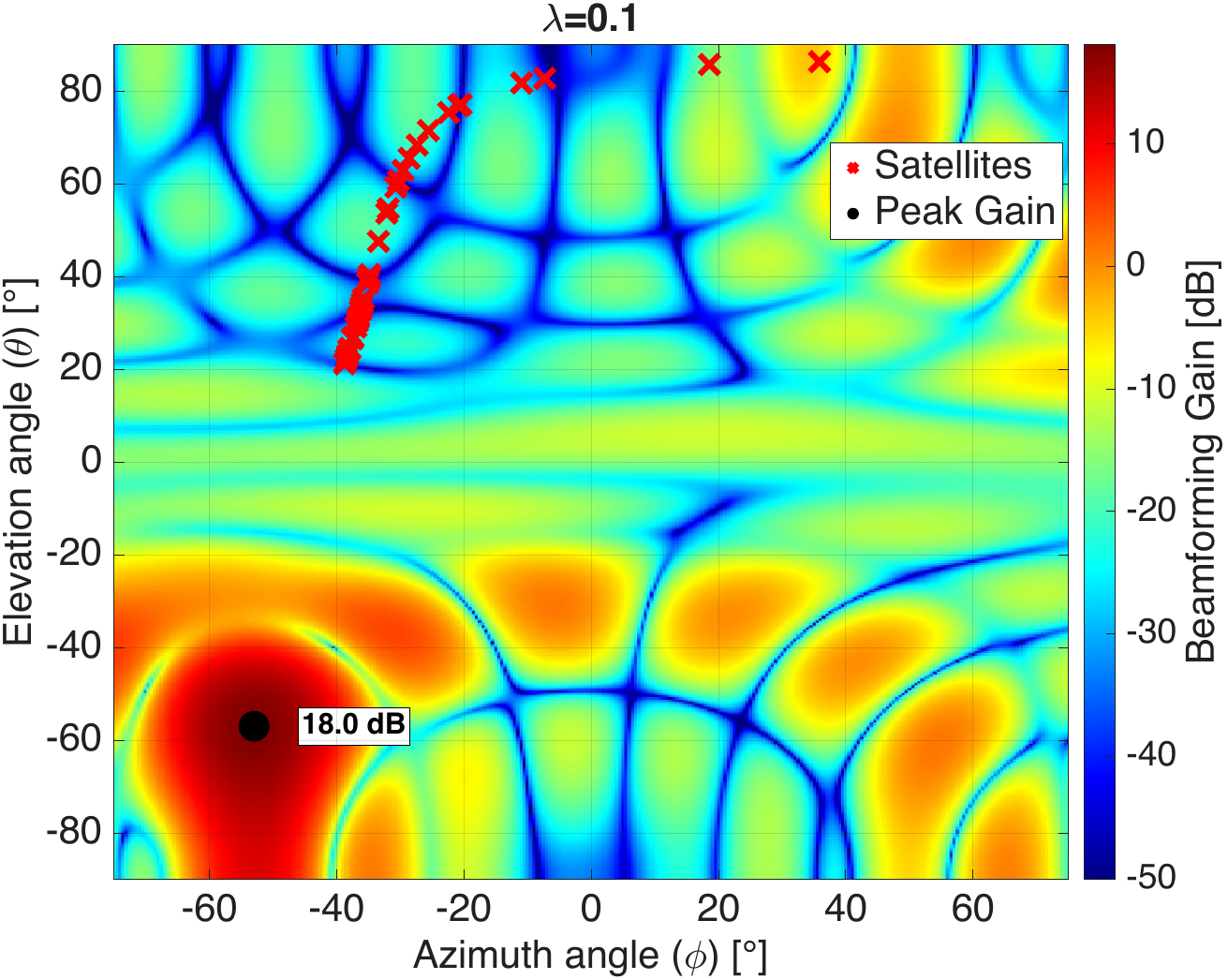}
\label{fig:bf-sim-start-01}}
%
%\hfil
%
\hspace{-0.25cm}
\subfigure[]{\includegraphics[height=3.5cm, width=4.4cm]{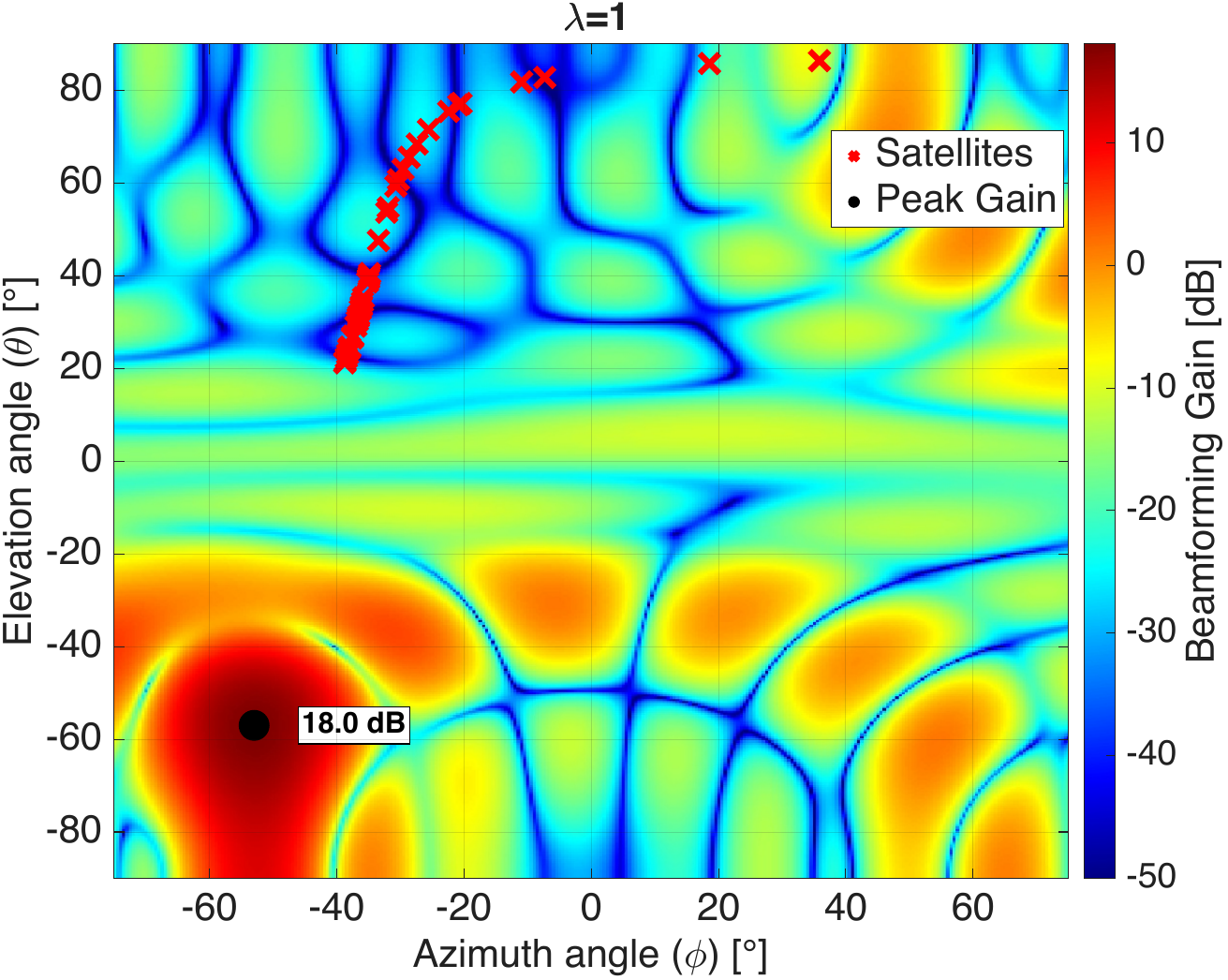}
\label{fig:bf-sim-start-1}}
%
%\hfil
%
\hspace{-0.22cm}
\subfigure[]{\includegraphics[height=3.5cm, width=4.4cm]{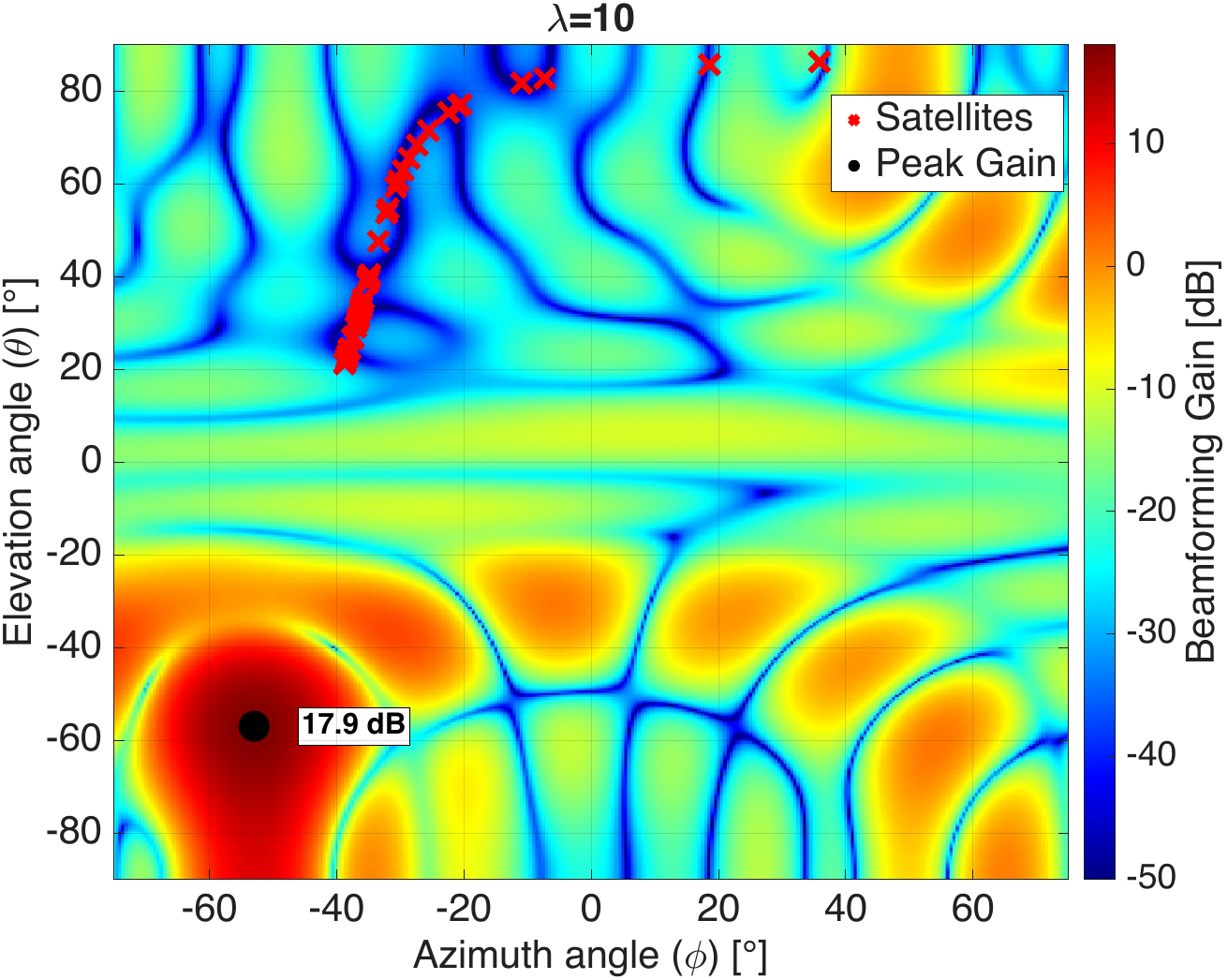}
\label{fig:bf-sim-start-10}}
\hspace{0.1\textwidth}
\setlength\abovecaptionskip{-.2cm}
% \caption{Beamforming gain at a terrestrial \gls{ue} at time instant $t_{l}$ along the satellites' trajectories, incurring minimal degradation to the terrestrial \gls{qos}.}
\caption{Beamforming gain of the considered \gls{gnb} with a $8\times 8$ \gls{upa} and $\lambda\in\{0,0.1,1,10\}$ for time instance ($t_{l}$) along the satellites' trajectories.
As $\lambda$ increases, a greater priority is given to the interference nulling than to the terrestrial network \gls{qos}, leading to deeper nulls toward the satellites and lower gain in the \gls{ue} direction.
However, the degradation is minimal: $18$~dB at $\lambda = 0$ (Fig.~\ref{fig:bf-sim-start-0}), to $17.9$~dB at $\lambda = 10$ (Fig.~\ref{fig:bf-sim-start-10}), close to the theoretical maximum of $18.06$~dB.}
\label{bf-sim-start}
\vspace{-0.5cm}
\end{figure*}

We begin by investigating and visually interpreting the Interference Nulling algorithm, examining the beamforming gain patterns. To show a more intuitive correspondence between the nulls and lobes and the physical locations of the satellites and the \glspl{ue}, respectively, in this subsection we run the algorithm considering a simplified \gls{los}-only channel instead of the \quadriga channel matrix by replacing $\tilde{\boldsymbol{h}}_j$ with $\boldsymbol{e}(\theta_j, \phi_j)$, where $(\theta_j, \phi_j)$ are the elevation and azimuth angles to the $j$-th victim satellite, and $\boldsymbol{e}(\theta_j, \phi_j)$ is the spatial signature of the \gls{gnb} antenna array in that direction. In addition, since satellite trajectories can be derived from publicly available ephemeris data, this approach can reduce the real-time complexity associated with multi-path channel estimation, while still achieving interference suppression close to that of full multi-path nulling, as reported in~\cite{kang2024terrestrial}.

Fig.~\ref{bf-sim-start} illustrates the resulting transmit beamforming gains $|\boldsymbol{e}^H(\theta, \phi)\boldsymbol{w}_t|^2$ across the angular space of a \gls{gnb} equipped with an $8\times 8$ \gls{upa}, as obtained by solving the interference nulling algorithm in~\eqref{eq:nulling-opt} for different values of $\lambda$ and a given satellite configuration. Specifically, the $N_{\text{sat}} = 40$ satellite locations are marked with red crosses, while the main beam direction (i.e., the peak gain) toward the terrestrial \gls{ue} is indicated by a black circle. The results qualitatively show that nulls can be effectively placed at all satellite locations while maintaining a high beamforming gain toward the desired terrestrial \gls{ue}. As $\lambda$ increases from $0$ to $10$, interference nulling is prioritized over terrestrial \gls{qos}, and deeper nulls appear at the satellite locations, providing stronger interference suppression. This comes at a minimal cost to the terrestrial \gls{ue} gain, which decreases from $18$~dB at $\lambda = 0$ (Fig.~\ref{fig:bf-sim-start-0}) to $17.9$~dB at $\lambda = 10$ (Fig.~\ref{fig:bf-sim-start-10}). Finally, the beamforming gain of $17.9$~dB remains close to the theoretical maximum of $10 \log_{10}(64) = 18.06$~dB, indicating minimal impact on terrestrial link quality.

\subsection{\gls{qos}-Aware Power Control for Risk-Averse Scenarios}\label{riskaverse}
As explained in Section~\ref{qos-pc}, the goal of the power control algorithm is to strike a balance between the interference toward the satellites and the terrestrial network \gls{qos}.

\begin{figure}[h!]
% \vspace{-1em}
\centering
\setlength\abovecaptionskip{-.22cm}
\includegraphics[width=\columnwidth]{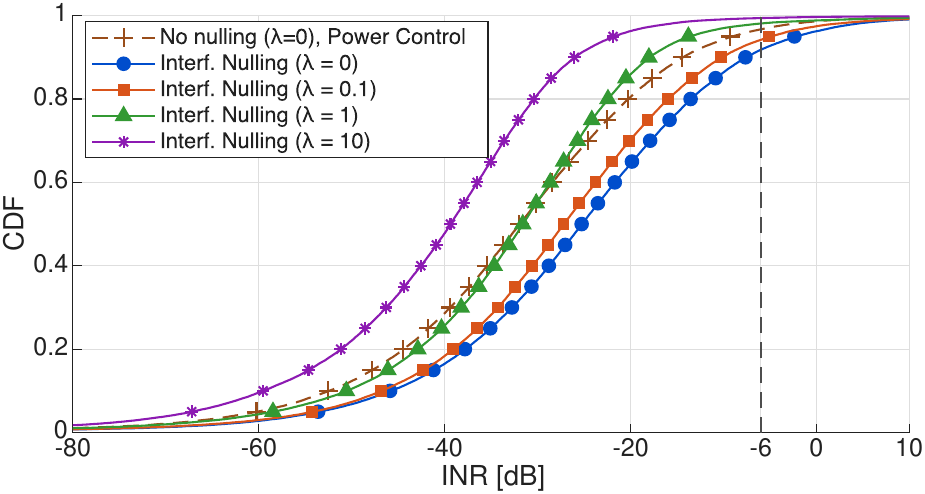}
\setlength\abovecaptionskip{-.1cm}
\caption{\gls{cdf} of the \gls{inr} as defined in~\eqref{eq:inr}: Power Control \& Standard Beamforming vs. Interference Nulling.}
\label{fig:inr-cdf-pc-null}
\vspace{-0.4cm}
\end{figure}

\begin{figure}[h!]
    \centering
    \hspace{-0.57cm}
    \begin{minipage}{0.32\columnwidth}
        \centering
        \input{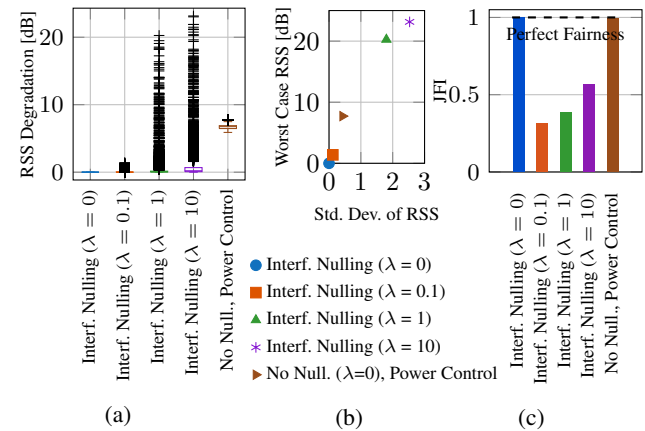}
        \centerline{\small (a)}
    \end{minipage}%
    % \hfill
    \hspace{0.15cm}
    \begin{minipage}{0.32\columnwidth}
        \centering
        \definecolor{mycolor1}{rgb}{0.06600,0.44300,0.74500}%
\definecolor{mycolor2}{rgb}{0.86600,0.32900,0.00000}%
\definecolor{mycolor3}{rgb}{0.52100,0.08600,0.81900}%
\definecolor{mycolor4}{rgb}{0.60000,0.30000,0.10000}%
\definecolor{mycolor5}{rgb}{0.12941,0.12941,0.12941}%
\begin{tikzpicture}
\begin{axis}[%
width=0.5in,
height=0.8in,
at={(0in,0in)},
scale only axis,
xmin=0,
xmax=3,
xlabel style={font=\scriptsize},
xlabel={Std. Dev. of RSS},
ymin=0,
ymax=25,
ylabel style={font=\scriptsize,yshift=-6pt},
ylabel={Worst Case RSS [dB]},
axis background/.style={fill=white},
xmajorgrids,
ymajorgrids,
legend style={at={(0.5,-0.54)}, anchor=north, legend columns=1, font=\scriptsize, legend cell align=left, draw=none}
]
\addplot[only marks, mark=*, mark options={}, mark size=2pt, draw=mycolor1, fill=mycolor1] table[row sep=crcr]{%
x	y\\
0	0\\
};
\addlegendentry{Interf. Nulling (λ = 0)}
\addplot[only marks, mark=square*, mark options={}, mark size=2pt, draw=mycolor2, fill=mycolor2] table[row sep=crcr]{%
x	y\\
0.119136109520964	1.38457900093323\\
};
\addlegendentry{Interf. Nulling (λ = 0.1)}
\addplot[only marks, mark=triangle*, mark options={}, mark size=2pt, draw=green!60!violet, fill=green!60!violet] table[row sep=crcr]{%
x	y\\
1.80113549433475	20.2563255543226\\
};
\addlegendentry{Interf. Nulling (λ = 1)}
\addplot[only marks, mark=asterisk, mark options={}, mark size=2pt, draw=mycolor3, fill=mycolor3] table[row sep=crcr]{%
x	y\\
2.53068651206873	23.1469268862214\\
};
\addlegendentry{Interf. Nulling (λ = 10)}
\addplot[only marks, mark=triangle*, mark options={rotate=270}, mark size=2pt, draw=mycolor4, fill=mycolor4] table[row sep=crcr]{%
x	y\\
0.425536896631504	7.72084358179715\\
};
\addlegendentry{No Null. (λ=0), Power Control}
\end{axis}
\end{tikzpicture}%
        \centerline{\small (b)}
    \end{minipage}%
    \hspace{-0.5cm}
    \begin{minipage}{0.32\columnwidth}
        \centering
        \definecolor{mycolor1}{rgb}{0.12941,0.12941,0.12941}%
\definecolor{bar1}{rgb}{0.0,0.30,0.80}% Deep Blue
\definecolor{bar2}{rgb}{0.85,0.33,0.10}% Strong Orange
\definecolor{bar3}{rgb}{0.20,0.60,0.20}% Rich Green
\definecolor{bar4}{rgb}{0.55,0.10,0.70}% Purple
\definecolor{bar5}{rgb}{0.6,0.3,0.1}% Brown
\begin{tikzpicture}[font=\small]
\begin{axis}[%
xtick style={draw=none},
width=0.8in,
height=0.85in,
ybar=0pt,
bar width=0.5,
scale only axis,
xmin=0.5,
xmax=5.5,
xtick={1,2,3,4,5},
xticklabels={{Interf. Nulling ($\lambda=0$)},{Interf. Nulling ($\lambda=0.1$)},{Interf. Nulling ($\lambda=1$)},{Interf. Nulling ($\lambda=10$)},{No Null., Power Control}},
xticklabel style={rotate=90, anchor=east, font=\scriptsize},
ymin=0,
ymax=1.05,
ylabel={JFI},
ylabel style={font=\scriptsize,yshift=-6pt},
axis background/.style={fill=white},
ymajorgrids,
clip=false,
enlarge x limits=0.15
]
\addplot[ybar, fill=bar1, draw=none, bar shift=0pt] coordinates {(1,1)};
\addplot[ybar, fill=bar2, draw=none, bar shift=0pt] coordinates {(2,0.313931050330797)};
\addplot[ybar, fill=bar3, draw=none, bar shift=0pt] coordinates {(3,0.384930621221513)};
\addplot[ybar, fill=bar4, draw=none, bar shift=0pt] coordinates {(4,0.568601454422722)};
\addplot[ybar, fill=bar5, draw=none, bar shift=0pt] coordinates {(5,0.995996633216235)};
\draw[black, dashed, thick, shorten >=2pt, shorten <=2pt] (axis cs:0.5,1) -- (axis cs:5.5,1) node[pos=0.5, below, font=\scriptsize] {Perfect Fairness};
\end{axis}
\end{tikzpicture}%
        \centerline{\small (c)}
    \end{minipage}
    \caption{Worst-case evaluation and fairness analysis for the terrestrial network topology described in Table~\ref{tab:sim_params}.}
    \label{fig:fairness}
    \vspace{-0.55cm}
\end{figure}

\noindent\textbf{\gls{inr} Reduction.} 
In Figs.~\ref{fig:inr-cdf-pc-null} and~\ref{fig:fairness}, we compare the simulation results obtained using the algorithm defined in~\eqref{eq:nulling-opt} with those obtained through the power control formulation defined in~\eqref{eq:power_control_obj}--\eqref{eq:power_control_power}.
It is noted that the terms \textit{Interference Nulling ($\lambda=0$)} and \textit{No Nulling} can be used interchangeably.
They both refer to the case where the trade-off between terrestrial \gls{qos} maximization and interference suppression—represented by the tunable parameter $\lambda$—is not considered, as explained in Section~\ref{interf-null}. In Fig.~\ref{fig:inr-cdf-pc-null}, we observe the \gls{inr} at the satellites for the simulation setup described in Table~\ref{tab:sim_params} and a throughput threshold of $\epsilon=0.85$.
We observe that Interference Nulling with $\lambda=10$ achieves the most interference reduction toward the satellites with a median \gls{inr} of $\sim40$~dB, while Interference Nulling with $\lambda=0.1$ performs almost identically to the case with Interference Nulling with $\lambda=0$ (equivalent to No Nulling). Similarly, Interference Nulling with $\lambda=1$ yields similar performance to Power Control ($\lambda=0$).
Finally, all methods successfully meet the \gls{itu} constraint by keeping the \gls{inr} at the satellites below $-6$~dB on average, with Interference Nulling for $\lambda \in \{1, 10\}$ and the combination of No Nulling and Power Control performing best.

% Although the \gls{inr} at the satellites can be mapped to terrestrial \gls{qos} degradation through Eq.~\eqref{eq:snr_degradation}, 
\noindent\textbf{\gls{rss} Degradation.} In Fig.~\ref{fig:fairness}, we analyze the \gls{ue} \gls{rss} degradation.
In Fig.~\ref{fig:fairness}(a), the simulation results indicate that although all Interference Nulling methods—regardless of the $\lambda$ parameter—achieve near-zero median \gls{rss} degradation on the terrestrial links, they exhibit extreme outliers, particularly as $\lambda$ increases. Conversely, Power Control combined with No Nulling results in a higher median \gls{rss} degradation, due to $P_i < P_{\rm max}$, but avoids the extreme outliers observed with $\lambda=1$ and $\lambda=10$. This approach maintains the worst-case degradation for all terrestrial links below $7$~dB, in contrast to the Interference Nulling methods, where the degradation can reach up to $\sim 20$~dB (Fig.~\ref{fig:fairness}(b)). This degradation indicates that \glspl{ue} at certain locations cannot be effectively served by the \gls{gnb} when using the interference nulling algorithm. 
To quantify this effect, we resort to the \gls{jfi}, reported in Fig.~\ref{fig:fairness}(c). We observe that \textit{Perfect Fairness} is achieved only for Interference Nulling with $\lambda=0$, whether performed standalone or combined with Power Control. In the former case, No Nulling represents the scenario where no \gls{qos} compromise is performed on the terrestrial network, resulting in a \gls{jfi} equal to unity. Similarly, the \gls{jfi} for the combined method is also equal to unity, as all terrestrial links are degraded equally. This can be attributed to the fact that Power Control reduces the transmit power across all links. In contrast, Interference Nulling compromises the channel quality of each \gls{gnb}--\gls{ue} link individually, depending on its specific channel state. Consequently, certain \glspl{ue} experience more severe \gls{qos} degradation, particularly as $\lambda$ increases to values of $1$ and $10$. Also, as $\lambda$ increases, a greater number of links are affected and experience similar levels of \gls{rss} degradation. Consequently, the \gls{jfi} also increases with $\lambda$; for instance, $\lambda=10$ results in more links with comparable and high \gls{qos} degradation compared to the degradation observed when $\lambda=0.1$ is used in the optimization. Therefore, except for the case when a simple \gls{svd} is performed toward the \gls{ue} ($\lambda=0$), the \gls{jfi} of the interference nulling method remains below $0.6$.

% \begin{figure}[h!]
%     \centering
%     \hspace{-1.5cm}
%     \begin{minipage}{0.39\columnwidth}
%         \centering
%         \includegraphics[width=\linewidth]{figures/Fairness_BoxPlot_DistributionSpread_AllUEs30_NoSats40_gNB1.pdf}
%         \centerline{\small \qquad (a)}
%         % \label{fig:fairness-box}
%     \end{minipage}%
%      \hspace{-0.1cm}
%     \begin{minipage}{0.48\columnwidth}
%         \centering
%         \includegraphics[width=\linewidth]{figures/Fairness_Scatter_WorstCaseVsVariability_AllUEs30_NoSats40_gNB1.pdf}
%         \centerline{\small (b)}
%         % \label{fig:fairness-scatter}
%     \end{minipage}%
%     \hspace{-1.5cm}
%     \begin{minipage}{0.35\columnwidth}
%         \centering
%         \includegraphics[width=\linewidth]{figures/Fairness_BarChart_JainsIndex_AllUEs30_NoSats40_gNB1.pdf}
%         \centerline{\small \qquad (c)}
%         % \label{fig:fairness-jfi}
%     \end{minipage}
%     \caption{Worst-case evaluation and fairness analysis for the terrestrial network topology described in Table~\ref{tab:sim_params}.}
%     \label{fig:fairness}
%     % \vspace{-0.3cm}
% \end{figure}

The aforementioned simulation results indicate that Interference Nulling with $\lambda=\{1, 10\}$, as well as No Nulling combined with Power Control, can keep the \gls{inr} at the satellites well under the \gls{itu} thresholds on average, thereby successfully minimizing interference at the satellites. However, Power Control is more suitable for risk-averse scenarios, since it manages to maintain an equal degradation for all links while avoiding extreme outliers for specific \glspl{ue}. Such an approach is well-suited for tactical or public safety applications, where the goal is to guarantee a reliable link for all \glspl{ue}, whereas Interference Nulling maintains high overall performance at the cost of compromising certain links to achieve a high average.

%%%%%%%%%%%%% TEX FILES FOR EDITING %%%%%%
% \begin{figure}[htbp]
%     \centering
%     \hspace{-0.05\columnwidth}
%     \begin{minipage}{0.38\columnwidth}
%         \centering
%         \input{figures/Fairness_BoxPlot_DistributionSpread_AllUEs30_NoSats40_gNB1.tex}
%         \centerline{\small \qquad\qquad  (a)}
%     \end{minipage}
%     \hfill
%     \begin{minipage}{0.28\columnwidth}
%         \centering
%         \input{figures/Fairness_Scatter_WorstCaseVsVariability_AllUEs30_NoSats40_gNB1.tex}
%         \centerline{\small \qquad (b)}
%     \end{minipage}
%     \hfill
%     \begin{minipage}{0.28\columnwidth}
%         \centering
%         \input{figures/Fairness_BarChart_JainsIndex_AllUEs30_NoSats40_gNB1.tex}
%         \centerline{\small \quad\qquad\qquad (c)}
%     \end{minipage}
%     \caption{Fairness evaluation across 30 UEs with 40 satellites and 1 gNB}
%     \label{fig:fairness}
% \end{figure}

\begin{figure}[ht!]
\centering
\hspace{-0.065\columnwidth}
\subfigure[]{
    \includegraphics[width=0.33\columnwidth]{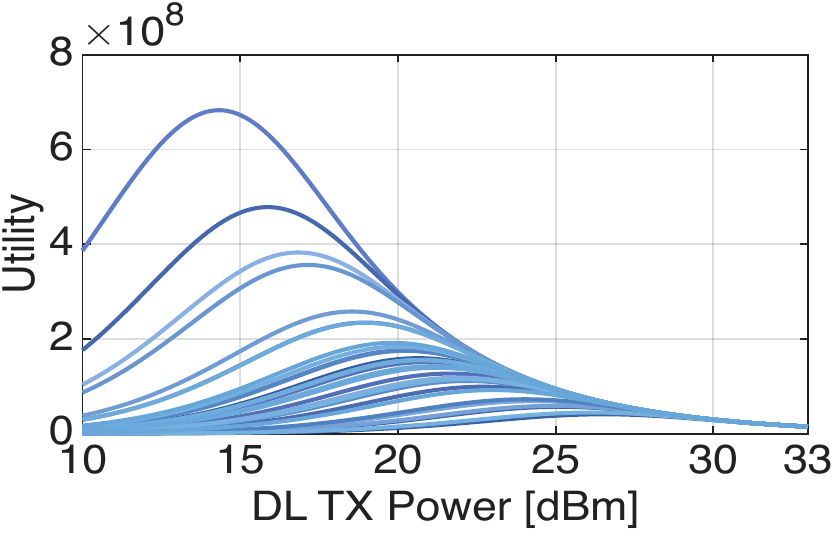}
    \label{fig:utility_all}
}
\hspace{-0.02\columnwidth}
\subfigure[]{
    \includegraphics[width=0.33\columnwidth]{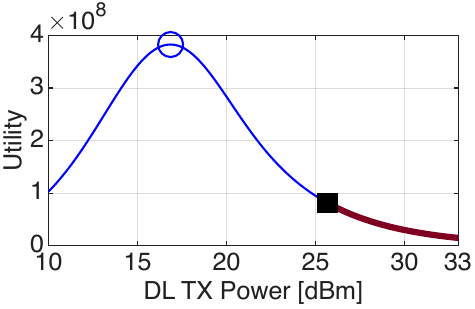}
    \label{fig:utility_ue11}
}
\hspace{-0.01\columnwidth}
\subfigure[]{
    \includegraphics[width=0.33\columnwidth]{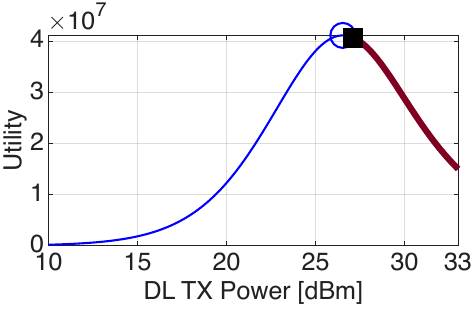}
    \label{fig:utility_ue9}
}
\caption{Utility function analysis for the terrestrial network topology described in Table~\ref{tab:sim_params}.}
\label{fig:utility}
% \vspace{-0.35cm}
\end{figure}

In Fig.~\ref{fig:utility}, the plots show the quasi-concavity, across the power levels considered, of the bell-shaped utility function formulated in~\eqref{eq:power_control_obj}. Fig.~\ref{fig:utility_all} showcases the optimal \gls{dl} transmit power chosen by the \gls{gnb} for each \gls{gnb}--\gls{ue} link, given the time-division-based scheduling decisions made by the \gls{gnb} scheduler for a power budget $[P_{\min}, P_{\max}] = [10, 33]$~dBm. Additionally, Fig.~\ref{fig:utility_all} displays the achieved utility per \gls{ue}, depicting the various individual \gls{qos} satisfaction levels. Indeed, the varying utility curves indicate that different power levels are selected for each active link, resulting in distinct utility peaks.
Figs.~\ref{fig:utility_ue11} and~\ref{fig:utility_ue9} demonstrate the selected power level versus the achieved utility at two distinct time instances. Both figures showcase, with a blue circle, the optimal power level that achieves the highest utility and \gls{qos} satisfaction (wrt. terrestrial \gls{sinr}) without unnecessary energy expenditures, as explained in detail in Section~\ref{qos-pc}. The shaded areas represent the power levels that satisfy the \gls{inr} and throughput constraints defined in~\eqref{eq:power_control_rate}--\eqref{eq:power_control_inr}, while the solid black square indicates the minimum power level that satisfies the throughput requirements while also meeting the \gls{itu} interference allowance constraints as selected by the power control formulation defined in~\eqref{eq:power_control_obj}--\eqref{eq:power_control_power}. It is noted that for the time instance represented by Fig.~\ref{fig:utility_ue9}, the optimal \gls{dl} transmit power level chosen by the \gls{gnb} closely matches the peak of the utility function. Conversely, for the time instance shown in Fig.~\ref{fig:utility_ue11}, the power level selected by the optimization is pushed to higher values to satisfy both the throughput and \gls{inr} constraints. Therefore, the utility function in this work inherently captures the trade-off between \gls{qos} satisfaction and unnecessary power expenditure. This allows the system to shift from energy-efficient operations to more \gls{qos}-driven regimes while still maintaining a balance in total energy consumption.

\begin{figure}[h!]
\centering
\hspace{-0.6cm}
\setlength\abovecaptionskip{-.15cm}
\subfigure[]{
    \scalebox{0.9}{% This file was created by matlab2tikz.
%
\definecolor{mycolor1}{rgb}{0.06600,0.44300,0.74500}%
\definecolor{mycolor2}{rgb}{0.12941,0.12941,0.12941}%
\begin{tikzpicture}

\begin{axis}[%
width=1.45in,
height=1in,
at={(0in,0in)},
scale only axis,
xmin=-66.7414407112592,
xmax=-54.5390548267721,
xlabel style={font=\footnotesize, yshift=3pt},
xlabel={Beamformed Channel Gain [dB]},
ymin=14.3261630815408,
ymax=26.5222611305653,
ylabel style={font=\footnotesize, yshift=-5pt},
ylabel={Utility Peak Power [dBm]},
axis background/.style={fill=white},
xmajorgrids,
ymajorgrids,
xtick distance=2,
ytick distance=3,
tick label style={font=\scriptsize},
]
\addplot[only marks, mark=*, mark options={}, mark size=0.75pt, color=mycolor1, fill=mycolor1, forget plot] table[row sep=crcr]{%
x	y\\
-54.5390548267721	14.3261630815408\\
-56.0853117830236	15.8679339669835\\
-57.0549047640633	16.8459229614807\\
-57.3642684368731	17.1450725362681\\
-58.7656648480108	18.5487743871936\\
-59.1773579986775	18.9629814907454\\
-59.1870662917992	18.9744872436218\\
-60.0527322308536	19.8374187093547\\
-60.2373958154455	20.0215107553777\\
-60.4246773111609	20.2056028014007\\
-60.4512180735895	20.24012006003\\
-60.8547619030881	20.6428214107054\\
-60.9480832308826	20.7348674337169\\
-60.9703486010627	20.7578789394697\\
-60.975643892087	20.7578789394697\\
-61.0879726891535	20.8729364682341\\
-61.3406787638822	21.1260630315158\\
-61.4417166629414	21.2296148074037\\
-61.8460283089614	21.632316158079\\
-62.1514299447453	21.9429714857429\\
-62.2807620095314	22.0695347673837\\
-62.4124910199673	22.1960980490245\\
-62.9119693359557	22.7023511755878\\
-63.2352739257249	23.0245122561281\\
-64.2614520537951	24.0485242621311\\
-64.5010760438395	24.2901450725363\\
-65.1763161067657	24.9574787393697\\
-65.331879238669	25.1185592796398\\
-66.3718723611911	26.1540770385193\\
-66.7414407112592	26.5222611305653\\
};
\end{axis}
\end{tikzpicture}%
}
    \label{fig:utility-peak}}
\hspace{-0.3cm}
\subfigure[]{
    \scalebox{0.9}{% This file was created by matlab2tikz.
%
\definecolor{mycolor1}{rgb}{0.06600,0.44300,0.74500}%
\definecolor{mycolor2}{rgb}{0.12941,0.12941,0.12941}%
\begin{tikzpicture}

\begin{axis}[%
width=1.45in,
height=1in,
at={(0in,0in)},
scale only axis,
xmin=-66.7414407112592,
xmax=-54.5390548267721,
xlabel style={font=\footnotesize, yshift=3pt},
xlabel={Beamformed Channel Gain [dB]},
ymin=0,
ymax=800000000,
ylabel style={font=\footnotesize, yshift=-5pt},
ylabel={Utility Peak},
axis background/.style={fill=white},
xmajorgrids,
ymajorgrids,
xtick distance=2,
tick label style={font=\scriptsize},
]
\addplot[only marks, mark=*, mark options={}, mark size=0.75pt, color=mycolor1, fill=mycolor1, forget plot] table[row sep=crcr]{%
x	y\\
-54.5390548267721	682984832.72717\\
-56.0853117830236	478393450.997838\\
-57.0549047640633	382671147.056436\\
-57.3642684368731	356360320.76784\\
-58.7656648480108	258077392.387138\\
-59.1773579986775	234736573.010031\\
-59.1870662917992	234212408.970654\\
-60.0527322308536	191886011.414123\\
-60.2373958154455	183897968.647588\\
-60.4246773111609	176136138.666808\\
-60.4512180735895	175063105.905671\\
-60.8547619030881	159529244.039796\\
-60.9480832308826	156137861.941875\\
-60.9703486010627	155339415.005414\\
-60.975643892087	155150070.061073\\
-61.0879726891535	151188681.152079\\
-61.3406787638822	142642414.127655\\
-61.4417166629414	139362150.902709\\
-61.8460283089614	126973686.430852\\
-62.1514299447453	118351323.948957\\
-62.2807620095314	114878894.162343\\
-62.4124910199673	111446743.140309\\
-62.9119693359557	99338900.1232685\\
-63.2352739257249	92212333.8895675\\
-64.2614520537951	72806698.4671001\\
-64.5010760438395	68898349.4275156\\
-65.1763161067657	58977264.4658116\\
-65.331879238669	56902143.9077569\\
-66.3718723611911	44784643.9598756\\
-66.7414407112592	41131269.6978792\\
};
\end{axis}
\end{tikzpicture}%
}
    \label{fig:bell-peak}}
\caption{Achieved utility and transmit power selection for the \gls{gnb} \gls{dl} under varying terrestrial channel conditions.}
\label{fig:utility-analysis}
\vspace{-0.25cm}
\end{figure}
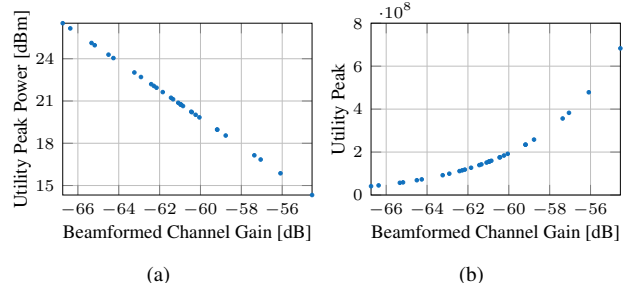

Fig.~\ref{fig:utility-analysis} shows the \gls{gnb}'s \gls{dl} transmit power and the associated achieved utility relative to the beamformed channel gain (i.e., $|\mathbf{w}_r^{\mathsf{H}} \tilde{\mathbf{H}}_{\text{ter}} \mathbf{w}_t|^2$) of the \gls{gnb}--\gls{ue} links.
% In Fig.~\ref{fig:utility-analysis}, we present the simulation results pertaining to the selection of the \gls{gnb}'s \gls{dl} transmit power and the associated achieved utility relative to the beamformed channel gain (i.e., $|\mathbf{w}_r^{\mathsf{H}} \tilde{\mathbf{H}}_{\text{ter}} \mathbf{w}_t|^2$) of the \gls{gnb}--\gls{ue} links. 
These results are obtained for the case where only \gls{svd} is employed on $\mathbf{H}_{\text{ter}}$, which is equivalent to interference nulling with $\lambda=0$. We observe that as the beamformed channel gain improves, the selected transmit power for the \gls{gnb}--\gls{ue} link decreases (Fig.~\ref{fig:utility-peak}), while the achieved utility—and consequently the \gls{qos} satisfaction of the link—increases (Fig.~\ref{fig:bell-peak}). Therefore, the power control framework proposed in this work allows the \gls{gnb} to select lower transmit power levels for terrestrial links with higher channel gains on the \gls{dl}, thereby avoiding excess interference toward the incumbents.

\subsection{Joint \gls{qos}-Aware Power Control and Interference Nulling}\label{combined-spectrum}
% \subsection{Joint \gls{qos}-Aware Power Control and Interference Nulling in Risk-Averse Scenarios}\label{combined-spectrum}
In this section, we present the simulation results when both Interference Nulling and Power Control are jointly considered and optimized to select the terrestrial beamforming weights and the \gls{gnb}'s \gls{dl} transmit power as summarized in Fig.~\ref{fig:flowchart}. We aim to explore how Interference Nulling with a less aggressive parameterization (i.e., lower $\lambda$ values) can be combined with Power Control to effectively minimize the \gls{inr} at the satellites while maintaining an acceptable terrestrial \gls{qos}.

\begin{figure}[h!]
  % \vspace{-1em}
  \centering
  \includegraphics[width=\columnwidth]
{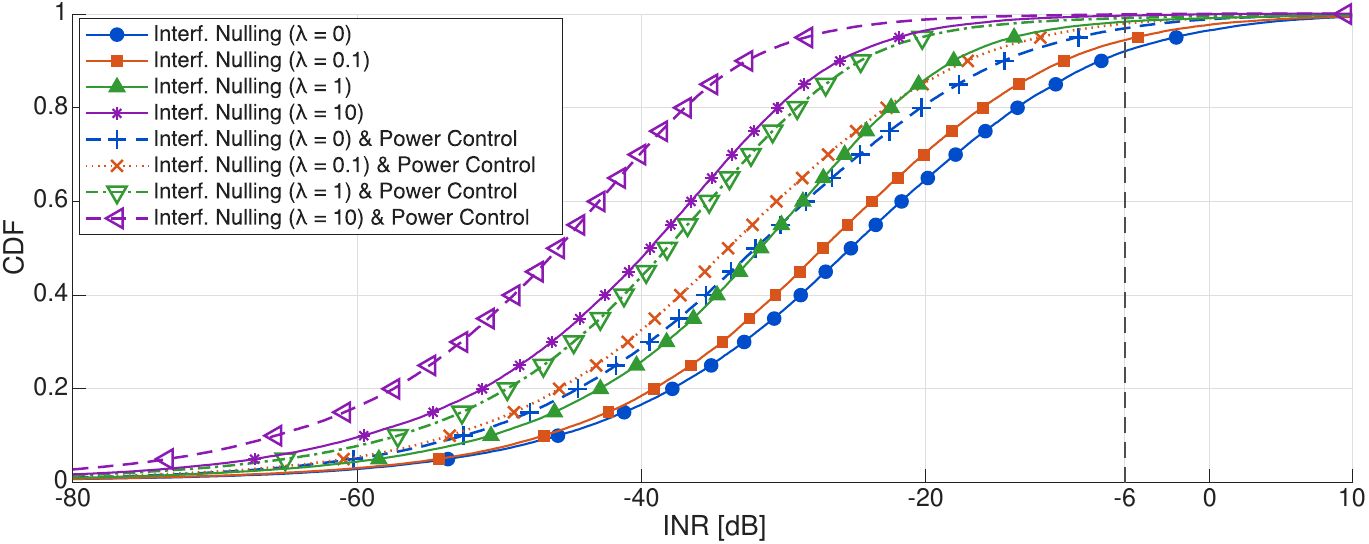} 
  \setlength\abovecaptionskip{-.12cm}
  % \vspace{-0.3cm}
  \caption{\gls{cdf} of the \gls{inr} as defined in~\eqref{eq:inr} for all optimization strategies.}
  \label{fig:inr-cdf-all}
  \vspace{-0.7cm}
\end{figure}

\noindent\textbf{\gls{inr} Reduction.} In Fig.~\ref{fig:inr-cdf-all}, we present the simulation results and the effect on the \gls{inr} at the satellites for all the methods considered in this work, namely Interference Nulling, No Nulling with Power Control, and the case of joint optimization through Interference Nulling and \gls{qos}-Aware Power Control. We observe that the most significant interference reduction towards the satellites is achieved through Interference Nulling with $\lambda=10$ combined with Power Control at a median \gls{inr} of $\sim46$~dB, while the least interference suppression is observed with Interference Nulling for $\lambda \in \{0, 0.1\}$. The most aggressive nulling technique observed with Interference Nulling ($\lambda=10$) yields similar performance to Interference Nulling with $\lambda=1$ combined with Power Control at a median \gls{inr} of $\sim38$~dB. Similarly, No Nulling combined with Power Control results in identical performance to Interference Nulling with $\lambda=1$. Therefore, when both power and the terrestrial beamformers are optimized jointly, the interference reduction at the satellites can be achieved with less aggressive nulling techniques (i.e., lower $\lambda$ values) given the proper adjustment of the \gls{gnb}'s \gls{dl} transmit power. Finally, \textit{Interf. Nulling ($\lambda=0.1$)} combined with Power Control provides a balanced trade-off, striking a middle ground between the less and more aggressive interference nulling techniques, by successfully keeping the interference leakage towards the satellites below the \gls{itu} interference levels.

\begin{figure}[h!]
  \vspace{-1em}
  \centering
  \includegraphics[width=0.65\columnwidth]
  {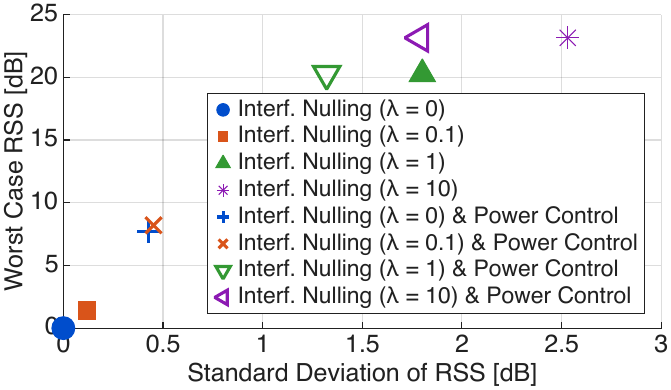} 
  \setlength\abovecaptionskip{-.1cm}
  % \vspace{-0.3cm}
  \caption{Worst-case \gls{rss} degradation as defined in~\eqref{eq:rss_degradation} for all optimization strategies.}
  \label{fig:worst-case-all}
  \vspace{-0.3cm}
\end{figure}

\noindent\textbf{\gls{rss} Degradation.} In Fig.~\ref{fig:worst-case-all}, we present the worst-case analysis results for all the methods. It is observed that Interference Nulling with $\lambda \in \{1, 10\}$, both standalone and when combined with Power Control, always results in the worst \gls{rss} values reported in our work; however, the joint optimization of the beamformers and the power results in more uniform \gls{rss} degradation across the terrestrial links (e.g., lower standard deviation). Additionally, Interference Nulling with $\lambda \in \{0, 0.1\}$ results in minimal, near-zero \gls{rss} degradation. Finally, No Nulling combined with Power Control achieves the same worst-case \gls{rss} as Interference Nulling with $\lambda=0.1$ also combined with Power Control, with the latter, however, achieving the best interference reduction among the two (Fig.~\ref{fig:inr-cdf-all}).

% Finally, to study the interference leakage from the \gls{gnb} toward the satellite \gls{ul}, we focus on lower elevation angles, as these correspond to scenarios with considerable terrestrial-to-satellite interference. Previous studies~\cite{li2002analytical,kang2024terrestrial} have shown that such angles may lead to increased interference due to the higher antenna gain of the \gls{gnb} in the direction of the satellite \gls{ul}. In addition, the probability of encountering satellites at lower elevation angles is higher, which represents a realistic use-case scenario. The probability density function (\gls{pdf}) of the satellite elevation angles observed from the considered constellation is shown in Fig.~\ref{fig:elevation-angle-pdf}.

% \begin{figure}[h!]
%     \centering
%     \setlength\abovecaptionskip{-.2cm}
%     \resizebox{\columnwidth}{!}{\input{figures/Elevation_Angle_PDF_gNB1_NoSats40.tex}}
%     \caption{\gls{pdf} of the satellite elevation angles for the constellation of $N_{\text{sat}} = 40$ satellites considered in this work.}
%     \label{fig:elevation-angle-pdf}
%     \vspace{-0.3cm}
% \end{figure}

\subsection{Combined \gls{qos}-Aware Power Control and Interference Nulling for Energy Efficiency Optimization}\label{combined-ee}

\begin{figure}[h!]
    \centering
    \hspace{-0.5cm}
    \begin{minipage}{0.35\columnwidth}
        \centering
        \resizebox{\linewidth}{!}{% This file was created by matlab2tikz.
% \documentclass[tikz]{standalone}
% \usepackage[T1]{fontenc}
% \usepackage[utf8]{inputenc}
% \usepackage{pgfplots}
% \usepackage{grffile}
% \pgfplotsset{compat=newest}
% \usetikzlibrary{plotmarks}
% \usetikzlibrary{arrows.meta}
% \usepgfplotslibrary{patchplots}
\usetikzlibrary{patterns}
\definecolor{mycolor1}{rgb}{0.06600,0.44300,0.74500}%
\definecolor{mycolor2}{rgb}{0.8500,0.3250,0.0980}%Orange
\definecolor{mycolor3}{rgb}{0.00,0.50,0.00}% 
\definecolor{mycolor4}{rgb}{0.60000,0.30000,0.10000}% Brown
\definecolor{mycolor5}{rgb}{0.95,0.00,0.00}% Very dark red

\definecolor{mycolorOutline}{rgb}{0.12941,0.12941,0.12941}%  % Dark gray for outline
\renewcommand{\Huge}{\fontsize{35}{42}\selectfont}
\begin{tikzpicture}
\begin{axis}[%
width=4.25in,
height=5in,
ytick distance=10,
at={(1.62in,2.45in)},
scale only axis,
xmin=0.2,
xmax=5.7,
xtick={1,2,3,4,5},
xticklabels={{Interf. Nulling ($\lambda$=0)},{Interf. Nulling ($\lambda$=0.1)},{Interf. Nulling ($\lambda$=1)},{Interf. Nulling ($\lambda$=0) \& Power Control ($\epsilon$=0.85)},{Interf. Nulling ($\lambda$=0.1) \& Power Control ($\epsilon$=0.98)}},
xticklabel style={rotate=90, anchor=east, font=\Huge},
yticklabel style={font=\Huge},
xlabel style={font=\Huge},
ymin=0,
ymax=35,
ylabel style={font=\Huge},
ylabel={Median INR [dB]},
axis background/.style={fill=white},
xmajorgrids,
ymajorgrids,
]
% Bar 1 - Blue with solid fill (no pattern)
\draw[fill=mycolor1, draw=mycolorOutline] (axis cs:0.675,0) rectangle (axis cs:1.325,25.1537084627834);
% Bar 2 - Orange with horizontal lines
\draw[pattern=horizontal lines, pattern color=mycolor2, draw=mycolorOutline] (axis cs:1.675,0) rectangle (axis cs:2.325,27.1641126299116);
% Bar 3 - Green-violet with vertical lines
\draw[pattern=vertical lines, pattern color=mycolor3, draw=mycolorOutline] (axis cs:2.675,0) rectangle (axis cs:3.325,31.811866293842);
% Bar 4 - Brown with north east lines (diagonal)
\draw[pattern=north east lines, pattern color=mycolor4, draw=mycolorOutline] (axis cs:3.675,0) rectangle (axis cs:4.325,31.8377824954678);
% Bar 5 - Sienna with crosshatch
\draw[pattern=crosshatch, pattern color=mycolor5, draw=mycolorOutline] (axis cs:4.675,0) rectangle (axis cs:5.325,28.0534984510458);
\node[above, align=center, inner sep=2pt, font=\huge]
at (axis cs:1,25.1537084627834) {-25.15};
\node[above, align=center, inner sep=2pt, font=\huge]
at (axis cs:2,27.1641126299116) {-27.16};
\node[above, align=center, inner sep=2pt, font=\huge]
at (axis cs:3,31.811866293842) {-31.81};
\node[above, align=center, inner sep=2pt, font=\huge]
at (axis cs:4,31.8377824954678) {-31.84};
\node[above, align=center, inner sep=2pt, font=\huge]
at (axis cs:5,28.0534984510458) {-28.05};
\addplot [color=black, dashed, line width=1.0pt, forget plot]
  table[row sep=crcr]{%
-0.225	0\\
6.225	0\\
};
\end{axis}
\end{tikzpicture}%
}
        \centerline{\small \qquad (a)}
    \end{minipage}%
    \hspace{0.2cm}
    \begin{minipage}{0.35\columnwidth}
        \centering
        \resizebox{\linewidth}{!}{\input{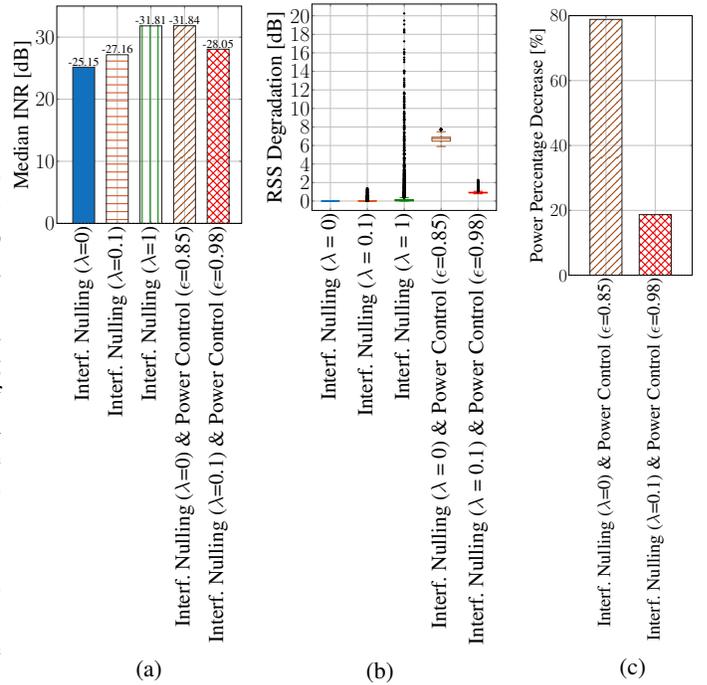}}
        \centerline{\small (b)}
    \end{minipage}%
    \hspace{0.3cm}
    \begin{minipage}{0.25\columnwidth}
        \centering
        \resizebox{\linewidth}{!}{\definecolor{mycolor2}{rgb}{0.12941,0.12941,0.12941}%
\definecolor{mycolor4}{rgb}{0.60000,0.30000,0.10000}% Brown
\definecolor{mycolor5}{rgb}{0.95,0.00,0.00}% Very dark red
\usetikzlibrary{patterns}
\renewcommand{\Huge}{\fontsize{35}{42}\selectfont}
\begin{tikzpicture}
\begin{axis}[%
width=3.25in,
height=6.85in,
ytick distance=20,
at={(1.625in,2.452in)},
scale only axis,
bar shift auto,
xmin=0.25,
xmax=2.75,
xtick={1,2},
xticklabels={{Interf. Nulling ($\lambda$=0) \& Power Control ($\epsilon$=0.85)},{Interf. Nulling ($\lambda$=0.1) \& Power Control ($\epsilon$=0.98)}},
xticklabel style={rotate=90, anchor=east, font=\Huge},
yticklabel style={font=\Huge},
xlabel style={font=\Huge},
ymin=0,
ymax=80,
ylabel style={font=\Huge},
ylabel={Power Percentage Decrease [\%]},
axis background/.style={fill=white},
xmajorgrids,
ymajorgrids
]
% Bar 1 - Brown with north east lines (diagonal)
\draw[pattern=north east lines, pattern color=mycolor4, draw=mycolor2] (axis cs:0.675,0) rectangle (axis cs:1.325,78.8519703093871);
% Bar 2 - Red with crosshatch
\draw[pattern=crosshatch, pattern color=mycolor5, draw=mycolor2] (axis cs:1.675,0) rectangle (axis cs:2.325,18.6858407834501);
\addplot[forget plot, color=mycolor2] table[row sep=crcr] {%
-0.225    0\\
3.225    0\\
};
\addplot [color=black, dashed, forget plot]
  table[row sep=crcr]{%
-0.225    0\\
3.225    0\\
};
\end{axis}
\end{tikzpicture}%
% \end{document}
}
        \centerline{\small \qquad (c)}
    \end{minipage}
    \caption{Performance evaluation of joint interference nulling and power control: (a) \gls{inr}, (b) \gls{rss} degradation, (c) power savings.}
    \label{fig:combined-plots}
    \vspace{-0.35cm}
\end{figure}

\begin{figure*}[h!]
    \centering
    \begin{minipage}{0.65\columnwidth}
        \centering
        \resizebox{\linewidth}{!}{% This file was created by matlab2tikz.
%
\definecolor{mycolor1}{rgb}{0.06600,0.44300,0.74500}%
\definecolor{mycolor2}{rgb}{0.86600,0.32900,0.00000}%
\definecolor{mycolor3}{rgb}{0.92900,0.69400,0.12500}%
\definecolor{mycolor4}{rgb}{0.52100,0.08600,0.81900}%
\definecolor{mycolor5}{rgb}{0.12941,0.12941,0.12941}%
\begin{tikzpicture}

\begin{axis}[%
width=8.88in,
height=3in,
at={(1.49in,0.738in)},
scale only axis,
xmin=0,
xmax=2.25,
xlabel={RSS Degradation [dB]},
ymin=0,
ymax=1,
ylabel={CDF},
axis background/.style={fill=white},
axis lines=box,
xmajorgrids,
ymajorgrids,
xtick distance=0.25,
ytick distance=0.2,
% Font sizes
% tick label style={font=\LARGE},
tick label style={font=\fontsize{30pt}{36pt}\selectfont},
xlabel style={font=\Huge},
ylabel style={font=\Huge},
legend style={font=\fontsize{28pt}{33pt}\selectfont, at={(0.97,0.03)}, anchor=south east, legend cell align=left, align=left}
]
\addplot [color=mycolor1, line width=4.0pt, mark=diamond*, mark size=6pt, mark repeat=3]
  table[row sep=crcr]{%
0.00185246207490491	0.0333333333333333\\
0.00509833290503491	0.0666666666666667\\
0.00809021092469777	0.1\\
0.0146075921803634	0.133333333333333\\
0.0160225575306208	0.166666666666667\\
0.0221507919148604	0.2\\
0.0319431163924386	0.233333333333333\\
0.0345364472297768	0.266666666666667\\
0.0347671416410916	0.3\\
0.0573645340633523	0.333333333333333\\
0.0825293959738525	0.366666666666667\\
0.113088431759504	0.4\\
0.114547951820576	0.433333333333333\\
0.144701775987775	0.466666666666667\\
0.158769357969512	0.5\\
0.230977863586902	0.533333333333333\\
0.293532237685226	0.566666666666667\\
0.302134751317143	0.6\\
0.344620881876663	0.633333333333333\\
0.350458822258672	0.666666666666667\\
0.565976394958641	0.7\\
0.581580907905105	0.733333333333333\\
0.582749627679227	0.766666666666667\\
0.709188839354131	0.8\\
0.784143570896785	0.833333333333333\\
0.864392213439205	0.866666666666667\\
0.885804064515884	0.9\\
1.54448563416248	0.933333333333333\\
2.07196611415236	0.966666666666667\\
2.13603830802803	1\\
};
\addlegendentry{$N_\text{t}=4\cdot4$}

\addplot [color=mycolor2, line width=4.0pt, mark=o, mark size=6pt, mark repeat=3]
  table[row sep=crcr]{%
6.40404585158867e-05	0.0333333333333333\\
0.000288495681511648	0.0666666666666667\\
0.000818528114426156	0.1\\
0.000862901919099855	0.133333333333333\\
0.000989036175149204	0.166666666666667\\
0.00101421100033435	0.2\\
0.00108147914424792	0.233333333333333\\
0.00111143922880703	0.266666666666667\\
0.0011833240804884	0.3\\
0.00134089297166161	0.333333333333333\\
0.00175659525129856	0.366666666666667\\
0.00223266171781281	0.4\\
0.0022655636365451	0.433333333333333\\
0.00331225459686888	0.466666666666667\\
0.00420602750156023	0.5\\
0.00566927972439358	0.533333333333333\\
0.00688299639290808	0.566666666666667\\
0.00865094193644288	0.6\\
0.0108205636699158	0.633333333333333\\
0.0123896889310632	0.666666666666667\\
0.023579116514832	0.7\\
0.0240075444170943	0.733333333333333\\
0.0344349230054297	0.766666666666667\\
0.0378774399232419	0.8\\
0.03968767678035	0.833333333333333\\
0.0918590485979865	0.866666666666667\\
0.206378893222341	0.9\\
0.293936616270972	0.933333333333333\\
1.18413294276332	0.966666666666667\\
1.61189597845284	1\\
};
\addlegendentry{$N_\text{t}=8\cdot8$}

\addplot [color=mycolor3, line width=4.0pt, mark=square*, mark size=6pt, mark repeat=3]
  table[row sep=crcr]{%
4.83186715542002e-06	0.0333333333333333\\
7.15451905509577e-06	0.0666666666666667\\
8.07984918448124e-06	0.1\\
8.45377660705411e-06	0.133333333333333\\
1.80876026339194e-05	0.166666666666667\\
0.000112193983191309	0.2\\
0.000126550352795987	0.233333333333333\\
0.000127111515119988	0.266666666666667\\
0.000141837284801798	0.3\\
0.000169173352690975	0.333333333333333\\
0.000178873454215494	0.366666666666667\\
0.000179176420447109	0.4\\
0.00018232685232262	0.433333333333333\\
0.000204637029501553	0.466666666666667\\
0.000413461381619056	0.5\\
0.00045873874214206	0.533333333333333\\
0.000538547244257568	0.566666666666667\\
0.000578965181708905	0.6\\
0.000748809661357579	0.633333333333333\\
0.000767112114162681	0.666666666666667\\
0.000801422919700636	0.7\\
0.000810577861862333	0.733333333333333\\
0.00464953053821828	0.766666666666667\\
0.00625083938742894	0.8\\
0.0089135723692456	0.833333333333333\\
0.0115159027884242	0.866666666666667\\
0.0149595151533708	0.9\\
0.0177762685689154	0.933333333333333\\
0.331838926262096	0.966666666666667\\
0.868238749945629	1\\
};
\addlegendentry{$N_\text{t}=16\cdot16$}

\addplot [color=mycolor4, line width=4.0pt, mark=x, mark size=6pt, mark repeat=3]
  table[row sep=crcr]{%
2.0708128085789e-06	0.0333333333333333\\
2.88999527562197e-06	0.0666666666666667\\
3.84238212270627e-06	0.1\\
7.29092175551612e-06	0.133333333333333\\
7.62676942353363e-06	0.166666666666667\\
8.29067612927337e-06	0.2\\
8.58479951520024e-06	0.233333333333333\\
9.50307013301094e-06	0.266666666666667\\
1.02855259547355e-05	0.3\\
2.05461723721586e-05	0.333333333333333\\
2.07409576262786e-05	0.366666666666667\\
2.10356916191024e-05	0.4\\
2.60770661809399e-05	0.433333333333333\\
2.76275140205225e-05	0.466666666666667\\
2.82479329229413e-05	0.5\\
2.8749855124729e-05	0.533333333333333\\
3.06723190623979e-05	0.566666666666667\\
5.9056737232768e-05	0.6\\
7.53634905244553e-05	0.633333333333333\\
0.000111811807117766	0.666666666666667\\
0.000115298117781328	0.7\\
0.000117724686314078	0.733333333333333\\
0.000223232690936574	0.766666666666667\\
0.00026503858686534	0.8\\
0.000560141635132469	0.833333333333333\\
0.000971968296514627	0.866666666666667\\
0.0009827151714882	0.9\\
0.00105036974027681	0.933333333333333\\
0.00387157926862293	0.966666666666667\\
0.10575336679986	1\\
};
\addlegendentry{$N_\text{t}=32\cdot32$}

\end{axis}
\end{tikzpicture}%
}
        \centerline{\small (a) $N_{\text{sat}} = 2$}
    \end{minipage}%
    \hfill
    \begin{minipage}{0.65\columnwidth}
        \centering
        \resizebox{\linewidth}{!}{% This file was created by matlab2tikz.
%
\definecolor{mycolor1}{rgb}{0.06600,0.44300,0.74500}%
\definecolor{mycolor2}{rgb}{0.86600,0.32900,0.00000}%
\definecolor{mycolor3}{rgb}{0.92900,0.69400,0.12500}%
\definecolor{mycolor4}{rgb}{0.52100,0.08600,0.81900}%
\definecolor{mycolor5}{rgb}{0.12941,0.12941,0.12941}%
\begin{tikzpicture}

\begin{axis}[%
width=8.88in,
height=3in,
at={(1.49in,0.738in)},
scale only axis,
xmin=0,
xmax=10,
xlabel={RSS Degradation [dB]},
ymin=0,
ymax=1,
ylabel={CDF},
axis background/.style={fill=white},
axis lines=box,
xmajorgrids,
ymajorgrids,
xtick distance=1,
ytick distance=0.2,
% Font sizes
% tick label style={font=\LARGE},
tick label style={font=\fontsize{30pt}{36pt}\selectfont},
xlabel style={font=\Huge},
ylabel style={font=\Huge},
legend style={font=\fontsize{28pt}{33pt}\selectfont, at={(0.97,0.03)}, anchor=south east, legend cell align=left, align=left}
]
\addplot [color=mycolor1, line width=4.0pt, mark=diamond*, mark size=6pt, mark repeat=3]
  table[row sep=crcr]{%
0.0611440028466047	0.0333333333333333\\
0.100743603567605	0.0666666666666667\\
0.14534769003968	0.1\\
0.201912315325306	0.133333333333333\\
0.258920696658482	0.166666666666667\\
0.259483663777716	0.2\\
0.307993632588901	0.233333333333333\\
0.331420025696819	0.266666666666667\\
0.365219562586999	0.3\\
0.36900418818323	0.333333333333333\\
0.423691322314855	0.366666666666667\\
0.425662052392153	0.4\\
0.514900022029413	0.433333333333333\\
0.737490770618864	0.466666666666667\\
0.778060699456951	0.5\\
0.792184639770753	0.533333333333333\\
0.812882512734538	0.566666666666667\\
0.851820888814258	0.6\\
0.870269136902936	0.633333333333333\\
0.897718754205092	0.666666666666667\\
1.41036138623223	0.7\\
1.64317377297736	0.733333333333333\\
1.70150220128834	0.766666666666667\\
1.97316351038906	0.8\\
2.27037694445967	0.833333333333333\\
2.67528871416514	0.866666666666667\\
2.84343559614091	0.9\\
7.67488843263015	0.933333333333333\\
9.42089859911793	0.966666666666667\\
9.80084229387366	1\\
};
\addlegendentry{$N_\text{t}=4\cdot4$}

\addplot [color=mycolor2, line width=4.0pt, mark=o, mark size=6pt, mark repeat=3]
  table[row sep=crcr]{%
0.00199443846603832	0.0333333333333333\\
0.00275707072813541	0.0666666666666667\\
0.00643708539460161	0.1\\
0.00667511643362909	0.133333333333333\\
0.00861901708103901	0.166666666666667\\
0.0095643898564886	0.2\\
0.0119733279778033	0.233333333333333\\
0.013962290220192	0.266666666666667\\
0.0168266379426772	0.3\\
0.021640685517219	0.333333333333333\\
0.0286467396706449	0.366666666666667\\
0.0438161843529442	0.4\\
0.0472933171248899	0.433333333333333\\
0.0537858878712069	0.466666666666667\\
0.056543150222863	0.5\\
0.0638151494935463	0.533333333333333\\
0.0660087405787577	0.566666666666667\\
0.0704007964711532	0.6\\
0.0781736128380582	0.633333333333333\\
0.0797576951823083	0.666666666666667\\
0.0812009973875893	0.7\\
0.0900530781377855	0.733333333333333\\
0.103750521935142	0.766666666666667\\
0.175852416937464	0.8\\
0.22483002423008	0.833333333333333\\
0.887701368052659	0.866666666666667\\
0.989062753346291	0.9\\
2.23242487879653	0.933333333333333\\
3.50902428301262	0.966666666666667\\
5.84368538391629	1\\
};
\addlegendentry{$N_\text{t}=8\cdot8$}

\addplot [color=mycolor3, line width=4.0pt, mark=square*, mark size=6pt, mark repeat=3]
  table[row sep=crcr]{%
0.000378498824302296	0.0333333333333333\\
0.00037925857025216	0.0666666666666667\\
0.00043884349511449	0.1\\
0.000546481070850689	0.133333333333333\\
0.000878565368150282	0.166666666666667\\
0.000918662965218198	0.2\\
0.001072092466995	0.233333333333333\\
0.00131667862896512	0.266666666666667\\
0.0016585418203277	0.3\\
0.00292325662994008	0.333333333333333\\
0.00366619646133799	0.366666666666667\\
0.00372669288522728	0.4\\
0.0037949628634118	0.433333333333333\\
0.00616204000413708	0.466666666666667\\
0.00709423047048688	0.5\\
0.00713094518483374	0.533333333333333\\
0.0100348228124021	0.566666666666667\\
0.0101234311900763	0.6\\
0.0145516793360326	0.633333333333333\\
0.0146485610069104	0.666666666666667\\
0.0175285610731069	0.7\\
0.0180778173849741	0.733333333333333\\
0.020346534853278	0.766666666666667\\
0.0344155902313644	0.8\\
0.0436970081105299	0.833333333333333\\
0.182408646368562	0.866666666666667\\
0.368193715397182	0.9\\
0.41465524712271	0.933333333333333\\
0.750227826678967	0.966666666666667\\
2.39346790144014	1\\
};
\addlegendentry{$N_\text{t}=16\cdot16$}

\addplot [color=mycolor4, line width=4.0pt, mark=x, mark size=6pt, mark repeat=3]
  table[row sep=crcr]{%
5.69285969134685e-05	0.0333333333333333\\
6.90112046332523e-05	0.0666666666666667\\
8.38726928855117e-05	0.1\\
0.000103824500398719	0.133333333333333\\
0.000146583067445744	0.166666666666667\\
0.000190573115701482	0.2\\
0.000205073693601723	0.233333333333333\\
0.000223158472966334	0.266666666666667\\
0.000330771322748681	0.3\\
0.000392273520265973	0.333333333333333\\
0.000452217337699372	0.366666666666667\\
0.000473505833207249	0.4\\
0.00048646097524304	0.433333333333333\\
0.00077992477197946	0.466666666666667\\
0.00111082648766845	0.5\\
0.00161379041376248	0.533333333333333\\
0.00165250589060079	0.566666666666667\\
0.00165533771967061	0.6\\
0.00175420251123978	0.633333333333333\\
0.0027297170880593	0.666666666666667\\
0.00290520014237938	0.7\\
0.00436187296864286	0.733333333333333\\
0.0043799408905278	0.766666666666667\\
0.0138754402275154	0.8\\
0.0180755445517012	0.833333333333333\\
0.0339796439679897	0.866666666666667\\
0.109407136919732	0.9\\
0.115922096705749	0.933333333333333\\
0.119481553428364	0.966666666666667\\
0.256789911032274	1\\
};
\addlegendentry{$N_\text{t}=32\cdot32$}

\end{axis}
\end{tikzpicture}%
}
        \centerline{\small (b) $N_{\text{sat}} = 10$}
    \end{minipage}%
    \hfill
    \begin{minipage}{0.65\columnwidth}
        \centering
        \resizebox{\linewidth}{!}{% This file was created by matlab2tikz.
%
\definecolor{mycolor1}{rgb}{0.06600,0.44300,0.74500}%
\definecolor{mycolor2}{rgb}{0.86600,0.32900,0.00000}%
\definecolor{mycolor3}{rgb}{0.92900,0.69400,0.12500}%
\definecolor{mycolor4}{rgb}{0.52100,0.08600,0.81900}%
\definecolor{mycolor5}{rgb}{0.12941,0.12941,0.12941}%
\begin{tikzpicture}

\begin{axis}[%
width=8.88in,
height=3in,
at={(1.49in,0.738in)},
scale only axis,
xmin=0,
xmax=28,
xlabel={RSS Degradation [dB]},
ymin=0,
ymax=1,
ylabel={CDF},
axis background/.style={fill=white},
axis lines=box,
xmajorgrids,
ymajorgrids,
xtick distance=2,
ytick distance=0.2,
% Font sizes
% tick label style={font=\LARGE},
tick label style={font=\fontsize{30pt}{36pt}\selectfont},
legend style={font=\fontsize{28pt}{33pt}\selectfont, at={(0.97,0.03)}, anchor=south east, legend cell align=left, align=left},
xlabel style={font=\Huge},
ylabel style={font=\Huge},
]
\addplot [color=mycolor1, line width=4.0pt, mark=diamond*, mark size=6pt, mark repeat=3]
  table[row sep=crcr]{%
0.150784343088984	0.0333333333333333\\
0.176914073082185	0.0666666666666667\\
0.256672393079011	0.1\\
0.374520791798248	0.133333333333333\\
0.390850283009693	0.166666666666667\\
0.463928630917365	0.2\\
0.660972834173043	0.233333333333333\\
0.727498811943934	0.266666666666667\\
0.738859283319612	0.3\\
0.870364445131341	0.333333333333333\\
1.09073328647381	0.366666666666667\\
2.09841827396183	0.4\\
2.29449110556897	0.433333333333333\\
2.38004350254743	0.466666666666667\\
2.73820113813495	0.5\\
2.79227469357596	0.533333333333333\\
3.24902225239094	0.566666666666667\\
3.34243593878675	0.6\\
3.95889267995385	0.633333333333333\\
4.37178303680653	0.666666666666667\\
4.65550714980012	0.7\\
4.78421102063783	0.733333333333333\\
5.8023756506039	0.766666666666667\\
6.91029166425237	0.8\\
7.11520769985376	0.833333333333333\\
8.13239961598975	0.866666666666667\\
9.20437524880089	0.9\\
17.2825473626598	0.933333333333333\\
26.1080597363247	0.966666666666667\\
27.0195885482789	1\\
};
\addlegendentry{$N_\text{t}=4\cdot4$}

\addplot [color=mycolor2, line width=4.0pt, mark=o, mark size=6pt, mark repeat=3]
  table[row sep=crcr]{%
0.0353844029983243	0.0333333333333333\\
0.0359226627796493	0.0666666666666667\\
0.0735084649833161	0.1\\
0.0800493418378649	0.133333333333333\\
0.0900529690103972	0.166666666666667\\
0.0954740351454275	0.2\\
0.101035443987375	0.233333333333333\\
0.107031334285951	0.266666666666667\\
0.10846999536044	0.3\\
0.11223016205084	0.333333333333333\\
0.164194498084884	0.366666666666667\\
0.181789830779762	0.4\\
0.209124603070517	0.433333333333333\\
0.225588519906638	0.466666666666667\\
0.27284707711987	0.5\\
0.278875658928573	0.533333333333333\\
0.327672488149953	0.566666666666667\\
0.40546088276772	0.6\\
0.478903960388396	0.633333333333333\\
0.627838662487565	0.666666666666667\\
0.827276346293179	0.7\\
1.06208144768094	0.733333333333333\\
1.15299337946193	0.766666666666667\\
1.20098611676122	0.8\\
1.45237885182073	0.833333333333333\\
1.65177592229646	0.866666666666667\\
1.83524733582869	0.9\\
4.69713182714115	0.933333333333333\\
6.20347511361891	0.966666666666667\\
7.02786845283056	1\\
};
\addlegendentry{$N_\text{t}=8\cdot8$}

\addplot [color=mycolor3, line width=4.0pt, mark=square*, mark size=6pt, mark repeat=3]
  table[row sep=crcr]{%
0.00363246326951044	0.0333333333333333\\
0.00600739494783401	0.0666666666666667\\
0.00824580286201115	0.1\\
0.00852528028816245	0.133333333333333\\
0.0109564843783053	0.166666666666667\\
0.0148508753473851	0.2\\
0.0156075045102645	0.233333333333333\\
0.0164385381009032	0.266666666666667\\
0.019345562985881	0.3\\
0.0265246330259201	0.333333333333333\\
0.042757607495839	0.366666666666667\\
0.0430453267660803	0.4\\
0.0464891499235645	0.433333333333333\\
0.0486133723411174	0.466666666666667\\
0.0522213624640864	0.5\\
0.0545972765655792	0.533333333333333\\
0.0568562800707586	0.566666666666667\\
0.0703160428447785	0.6\\
0.0895500797894316	0.633333333333333\\
0.102088187939036	0.666666666666667\\
0.237054937535303	0.7\\
0.243200362593343	0.733333333333333\\
0.260910637355988	0.766666666666667\\
0.328477666441404	0.8\\
0.363070990237147	0.833333333333333\\
0.751462851012756	0.866666666666667\\
0.848642473196218	0.9\\
1.10487999618769	0.933333333333333\\
1.229093606269	0.966666666666667\\
2.57941551638496	1\\
};
\addlegendentry{$N_\text{t}=16\cdot16$}

\addplot [color=mycolor4, line width=4.0pt, mark=x, mark size=6pt, mark repeat=3]
  table[row sep=crcr]{%
0.00053604606142364	0.0333333333333333\\
0.00215064372876698	0.0666666666666667\\
0.00227824012899382	0.1\\
0.00230512435903118	0.133333333333333\\
0.00254223303965103	0.166666666666667\\
0.00269770036058914	0.2\\
0.00422339864310101	0.233333333333333\\
0.00448547062847651	0.266666666666667\\
0.00469700475910989	0.3\\
0.00797671810265893	0.333333333333333\\
0.00832471853217746	0.366666666666667\\
0.00923760990169438	0.4\\
0.0111108826964904	0.433333333333333\\
0.0117700120953546	0.466666666666667\\
0.0130320635061294	0.5\\
0.0137457542058332	0.533333333333333\\
0.0175182167026937	0.566666666666667\\
0.0185532357503033	0.6\\
0.0191846790322462	0.633333333333333\\
0.0213266323616321	0.666666666666667\\
0.0213282133482906	0.7\\
0.0234641454562958	0.733333333333333\\
0.0252851676630107	0.766666666666667\\
0.0311061385115086	0.8\\
0.0440242160841996	0.833333333333333\\
0.0498812081392195	0.866666666666667\\
0.290989445818881	0.9\\
0.299169440655581	0.933333333333333\\
0.426713421045725	0.966666666666667\\
0.500959309984685	1\\
};
\addlegendentry{$N_\text{t}=32\cdot32$}

\end{axis}
\end{tikzpicture}%
}
        \centerline{\small (c) $N_{\text{sat}} = 40$}
    \end{minipage}
    \caption{\gls{cdf} of \gls{rss} degradation for $\lambda = 1$ with varying number of satellites.}
    \label{fig:rss-degradation-satellites}
    \vspace{-0.18cm}
\end{figure*}
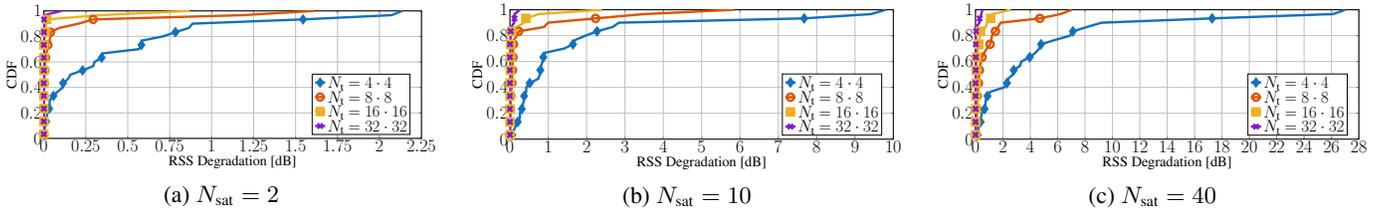

\begin{figure*}[h!]
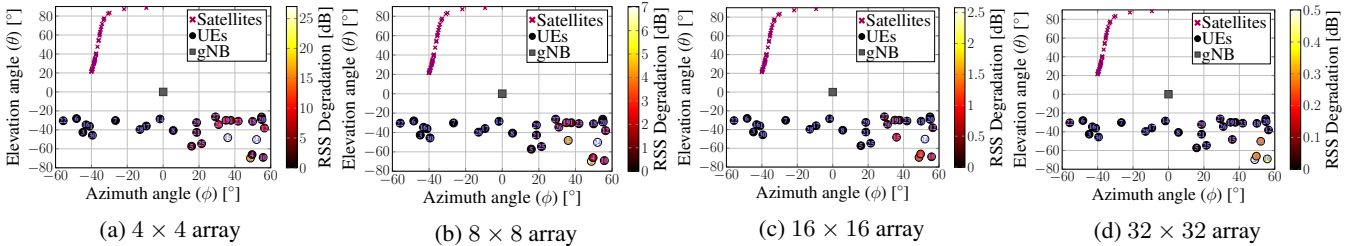

    \centering
    \vspace{-3.75pt}
    \setlength\abovecaptionskip{-.02cm}
    \begin{minipage}{0.5\columnwidth}
        \centering
        \resizebox{\linewidth}{!}{\input{figures/Fig7_AngularPositions_gNB1_NoSats40_Nt4x4_lambda1_snap1.tex}}
        \vspace{2pt}
        \centerline{\small (a) $4 \times 4$ array}
    \end{minipage}%
    \hspace{-1.5pt}
    \begin{minipage}{0.5\columnwidth}
        \centering
        \resizebox{\linewidth}{!}{\input{figures/Fig7_AngularPositions_gNB1_NoSats40_Nt8x8_lambda1_snap1.tex}}
        \vspace{2pt}
        \centerline{\small (b) $8 \times 8$ array}
    \end{minipage}%
    \hspace{-1.5pt}
    \begin{minipage}{0.5\columnwidth}
        \centering
        \resizebox{\linewidth}{!}{\input{figures/Fig7_AngularPositions_gNB1_NoSats40_Nt16x16_lambda1_snap1}}
        \vspace{2pt}
        \centerline{\small (c) $16 \times 16$ array}
    \end{minipage}%
    \hspace{-1.5pt}
    \begin{minipage}{0.5\columnwidth}
        \centering
        \resizebox{\linewidth}{!}{\input{figures/Fig7_AngularPositions_gNB1_NoSats40_Nt32x32_lambda1_snap1}}
        \vspace{2pt}
        \centerline{\small (d) $32 \times 32$ array}
    \end{minipage}
    \caption{Angular positions for the complete topology for different antenna sizes with $N_{\text{sat}} = 40$ satellites and Interference Nulling with $\lambda=1$.}
    \label{fig:angular-total_plot}
    \vspace{-0.5cm}
\end{figure*}

\begin{figure}[h!]
  \centering
  \subfigure[Avg. Satellite Channel Gain]{
    \includegraphics[width=0.96\columnwidth]{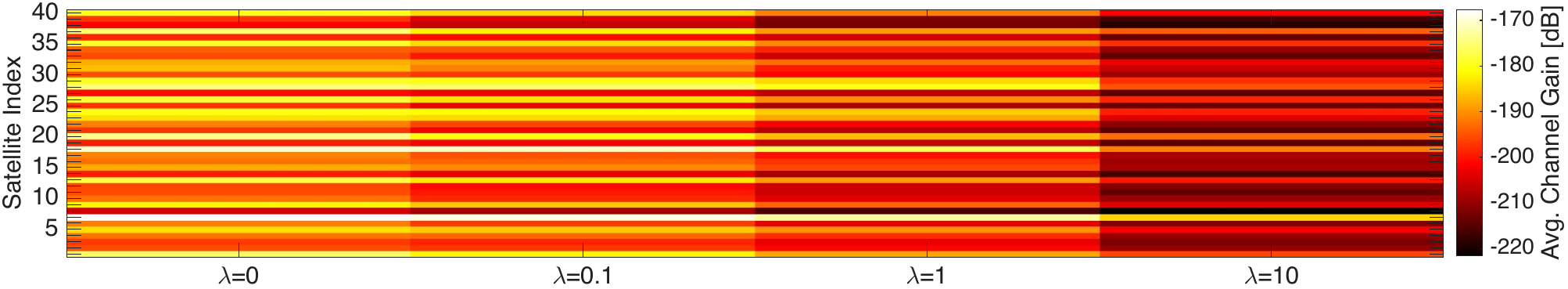}
    \label{fig:channel_strength}
  }
  \hspace{0.2\columnwidth}
  \subfigure[Terrestrial \gls{ue} \gls{rss} Degradation]{
    \includegraphics[width=0.96\columnwidth]{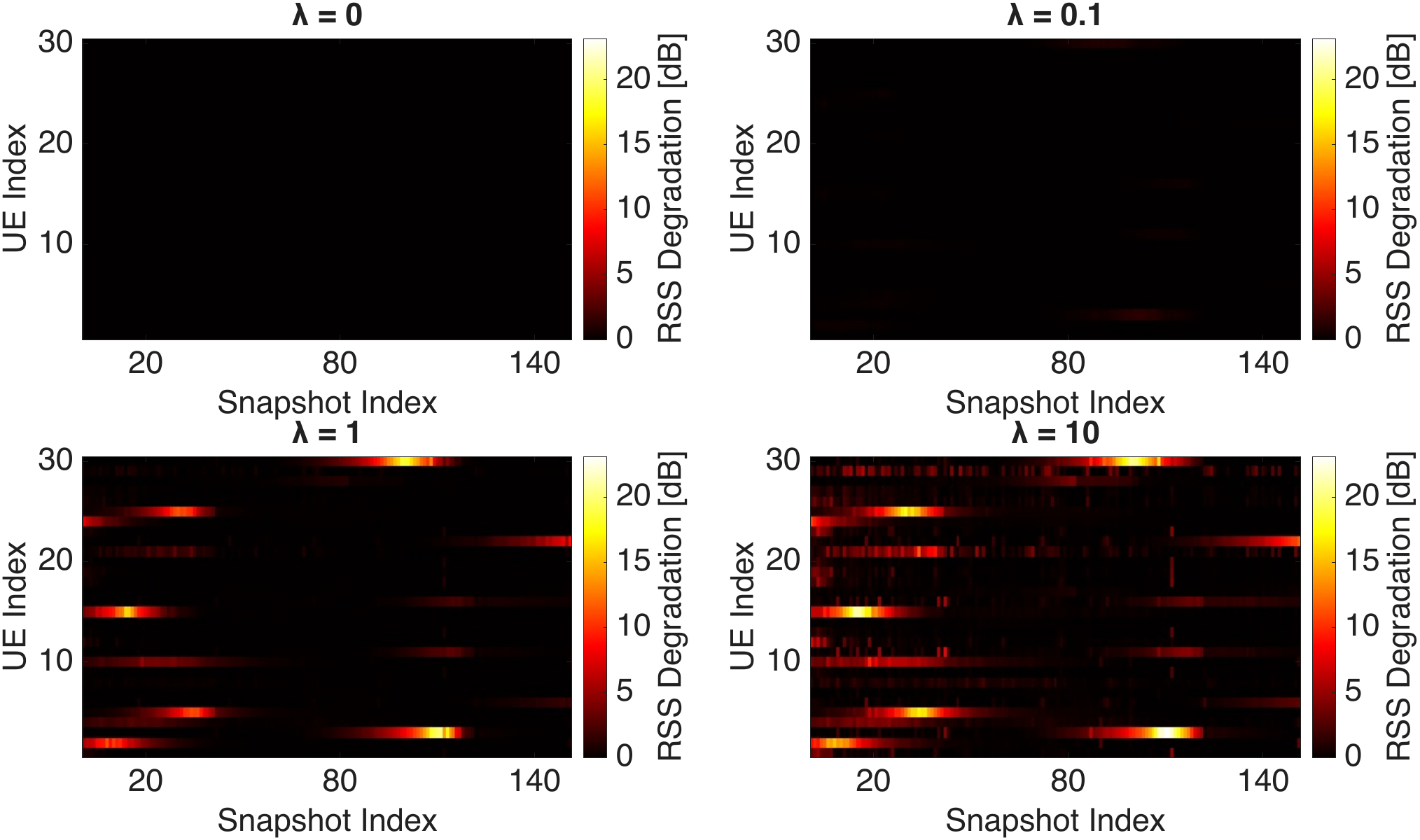}
    \label{fig:heatmap_deg}
  }
  \setlength\abovecaptionskip{-.1cm}
 \caption{Channel analysis: (a) Average signal leakage towards all satellites, and (b) \gls{ue} \gls{rss} degradation over satellite movement as defined in~\eqref{eq:rss_degradation} for all Interference Nulling techniques.}
  \label{fig:channel_analysis}
  \vspace{-0.75cm}
\end{figure}

\noindent\textbf{\gls{inr} Reduction, \gls{rss} Degradation and Power Savings.} In Fig.~\ref{fig:combined-plots}, we present the simulation results obtained when both Interference Nulling and Power Control are applied. We focus on the case with $\lambda = 1$, which was previously identified as providing the best trade-off between interference mitigation and acceptable \gls{qos} in the terrestrial network when only \textit{Interference Nulling} was considered. We additionally consider two different throughput thresholds (i.e., $\epsilon=\{0.85, 0.98\}$), as given in Table~\ref{tab:sim_params}, which represent different thresholds for the \gls{qos} requirements in the terrestrial network, as explained in Section~\ref{qos-pc}. In Fig.~\ref{fig:combined-plots}(a), we observe that Interference Nulling with $\lambda=1$ yields identical performance to Interference Nulling with $\lambda=0$ when power control is applied with $\epsilon=0.85$, providing an $\sim 7$~dB decrease in the median interference leakage at the satellite compared to the baseline (i.e., only Interference Nulling with $\lambda = 0$). Additionally, Interference Nulling with $\lambda = 0.1$ combined with power control with $\epsilon = 0.98$ (i.e., only a $2\%$ loss in the \gls{dl} throughput) manages to reduce the \gls{inr} at the satellites by approximately $3$~dB, compared to Interference Nulling alone (i.e., $\lambda = 0.1$),
for which the corresponding interference reduction compared to the baseline (i.e., $\lambda = 0$) is about $2$~dB.
Likewise, as Fig.~\ref{fig:combined-plots}(b) demonstrates, Interference Nulling combined with power control manages to balance the degradation in the \gls{qos} across the \glspl{ue}. In particular, the worst \gls{rss} degradation is reported to be less than $8$~dB when power control with $\epsilon=0.85$ is considered, compared to worst-case values of about $20$~dB obtained with Interference Nulling with $\lambda=1$. In contrast, power control with $\epsilon = 0.98$ combined with Interference Nulling with $\lambda=0.1$ manages to maintain an almost uniform degradation of approximately $1$~dB across all \glspl{ue}, with the worst \gls{rss} degradation reported at $\sim2$~dB. Therefore, power control with a dedicated throughput threshold can be combined with lower $\lambda$ values to balance the \gls{qos} degradation among all \glspl{ue}, while suppressing the interference leakage toward the satellites and additionally providing energy savings (Fig.~\ref{fig:combined-plots}(c)). Indeed, Fig.~\ref{fig:combined-plots}(c) demonstrates the energy savings achieved with both of the aforementioned power control techniques. Intuitively, more aggressive power control (i.e., $\epsilon = 0.85$) manages to save up to $80\%$ of the energy at the \gls{gnb}, at the cost of a $6.6$~dB median degradation across all terrestrial \glspl{ue}. Interference Nulling with $\lambda = 0.1$ combined with less aggressive power control ($\epsilon = 0.98$), on the other hand, also manages to save up to $20\%$ of the energy at the terrestrial \gls{gnb}, providing the best trade-off among all considered techniques in terms of terrestrial \gls{qos} and interference suppression.

% \subsection{Limitations of Interference Nulling and the Case for Power Control}\label{nulling-limitations}
% \subsection{Motivating Power Control: Interference Nulling  Limitations }\label{nulling-limitations}
\subsection{Case for Power Control: Limitations of Interference Nulling}\label{nulling-limitations}

In Fig.~\ref{fig:rss-degradation-satellites}, we present the \gls{cdf} of the \gls{rss} degradation for various antenna array sizes and number of satellites, focusing on the Interference Nulling technique with $\lambda = 1$. We observe that for a small constellation size (e.g., $N_{\text{sat}} = 2$ in Fig.~\ref{fig:rss-degradation-satellites}(a)), Interference Nulling can effectively keep the terrestrial \gls{qos} degradation low for all considered antenna sizes. Specifically, when larger antenna arrays are employed (i.e., $N_t \in \{256, 1024\}$), the degradation is maintained below $1$ and $0.25$ dB, respectively. However, as the number of satellites increases to $10$ and $40$ (Figs.~\ref{fig:rss-degradation-satellites}(b) and \ref{fig:rss-degradation-satellites}(c)), larger antenna arrays are required to keep the \gls{rss} degradation at low levels. For instance, in the case of $40$ satellites, an array size of $N_t = 1024$ elements is necessary to maintain the degradation below $1$ dB (Fig.~\ref{fig:rss-degradation-satellites}(c)). The simulation results demonstrate that antenna arrays with a higher number of elements yield more numerous and precise beams. This improved terrestrial performance aligns with the fundamental relationship \gls{dof} $= N_t - 1$, wherein increased element count provides additional \gls{dof} for null steering without compromising main lobe gain~\cite{balanis2016antenna,van2002optimum}.

Our analysis in Fig.~\ref{fig:angular-total_plot} further reveals that the degradation in terrestrial \gls{ue} performance due to Interference Nulling is primarily driven by the \gls{ue} angular position with respect to the \gls{gnb} \gls{upa} and the satellite position. Specifically, \glspl{ue} at 
positive azimuth angles (e.g., $\phi>0^\circ$), symmetric to the satellite positions, experience the highest degradation. Indeed, most affected \glspl{ue} are located at positive azimuth angles (e.g., $\phi=20^\circ$ to $60^\circ$) and negative elevations (e.g., $\theta<-20^\circ$), while satellites are located at negative azimuth angles (e.g., $\phi=-20^\circ$ to $-60^\circ$) and positive elevations (e.g., $\theta>20^\circ$). This represents a fundamental geometric trade-off inherent to null steering: constraining the beamformer to create nulls toward satellites at positive elevations inherently reduces the achievable gain for \glspl{ue} at extreme negative elevations, with the effect most pronounced for \glspl{ue} in the opposite azimuth quadrant. Importantly, this effect persists even with larger antenna arrays; while increasing the array size from $4\times4$ to $32\times32$ significantly reduces the degradation magnitude (from $25$~dB to $0.5$~dB), the same \glspl{ue} consistently exhibit the worst relative performance (e.g., $\theta<-40^\circ$ and $40^\circ<\phi<60^\circ$). As further detailed in \cite{wadaskar2025satellite}, the array response satisfies $\mathbf{a}(\theta_{az}, \theta_{el})=\mathbf{a}(-\theta_{az},-\theta_{el})$ since $\sin(-\theta_{az})\sin(-\theta_{el})=\sin(\theta_{az})\sin(\theta_{el})$ and $\cos(-\theta_{el})=\cos(\theta_{el})$. This symmetry implies that steering nulls toward satellites at $(\theta_{az}^{\text{sat}}, \theta_{el}^{\text{sat}})$ simultaneously affects beamforming gain at $(-\theta_{az}^{\text{sat}}, -\theta_{el}^{\text{sat}})$, limiting the \gls{dof} for terrestrial beamforming. These findings demonstrate that null steering introduces an unavoidable geometric coupling between the nulled and served directions, which can only be partially mitigated through increased array dimensions. 

In Fig.~\ref{fig:channel_analysis}, we present simulation results showing the effect of the Interference Nulling strategies (formulated in Section~\ref{interf-null}) on the channels for different $\lambda$ values. Fig.~\ref{fig:channel_analysis}(a) illustrates how the signal leakage towards the satellites (averaged across all snapshots of the satellite movement and, denoted as $|\tilde{\boldsymbol{h}}_j^{\text{H}} \boldsymbol{w}_t|$), is mitigated—particularly as $\lambda$ increases—demonstrating the effectiveness of the interference nulling techniques, especially for $\lambda$ values of $1$ and $10$. Fig.~\ref{fig:channel_analysis}(b) shows the effects of the interference nulling techniques on all \glspl{ue} throughout the satellite movement. Notably, $\lambda=0.1$ incurs minimal degradation on the terrestrial network across all satellite snapshot indices, whereas $\lambda=1$, and particularly $\lambda=10$, demonstrate a stronger impact on specific \glspl{ue}, which can experience up to $\sim 20$~dB \gls{rss} degradation. Therefore, Fig.~\ref{fig:channel_analysis} indicates that, although the average \gls{rss} degradation remains near zero, specific \glspl{ue} can experience severe \gls{qos} loss at certain points along the satellite trajectories, highlighting the limitations and significant trade-offs of Interference Nulling for satellite interference suppression while maintaining acceptable terrestrial \gls{qos}.

\section{Conclusions and Future Work}\label{conculsion}
We have performed a comparative analysis of two state-of-the-art techniques used for interference suppression to satellites, namely \gls{qos}-aware Power Control and Interference Nulling, in spectrum sharing scenarios for the \gls{fr3} band. Our simulation results demonstrated the suitability of power control techniques for risk-averse scenarios, while pinpointing the specific geometric and array-size limitations of interference nulling. Specifically, we demonstrated that interference nulling can severely degrade \glspl{ue} at certain angular positions; however, this degradation can be significantly mitigated by employing larger antenna array sizes. Finally, we proposed the joint application of interference nulling and power control techniques to reduce interference towards the incumbents while ensuring a more energy-efficient operation and fair \gls{qos} distribution in the terrestrial network. Our future work will focus on the on-demand and adaptive per-\gls{ue} optimization of power and beamforming. Specifically, we aim to apply power control selectively only for \glspl{ue} experiencing significant \gls{qos} degradation, thereby maintaining the \gls{rss} degradation at manageable levels while preserving system-wide performance. Another research focus will be the joint optimization of transmit power levels and beamforming for \gls{mu}-\gls{mimo} systems, where power allocation decisions can be made proportionally fair to the links' channel quality.

\vspace{-0.1cm}
%\balance
\bibliographystyle{IEEEtran}
\bibliography{IEEEabrv,ref}

@online{mgen,
    title = {{U.S. Naval Research Laboratory, "Multi-Generator ({MGEN}) Network Test Tool"}},
    howpublished={\url{https://www.nrl.navy.mil/itd/ncs/products/mgen}},
    note={2019}
}

@techreport{jaeckel2019f,
  title = "{F. Burkhardt, E. Eberlein,“QuaDRiGa-Quasi Deterministic Radio Channel Generator,” User Manual and Documentation}",
  author = {Jaeckel, S and Raschkowski, L and Boerner, K and Thiele, L},
  year = {2019},
  institution = {Tech. Rep. v2. 2.0, Fraunhofer Heinrich Hertz Institute},
  note = {[Online]. Available: \url{https://github.com/fraunhoferhhi/QuaDRiGa/blob/main/quadriga_documentation_v2.8.1-0.pdf}}
}

@article{jaeckel2014quadriga,
  title="{QuaDRiGa: A 3-D multi-cell channel model with time evolution for enabling virtual field trials}",
  author={Jaeckel, Stephan and Raschkowski, Leszek and B{\"o}rner, Kai and Thiele, Lars},
  journal={IEEE transactions on antennas and propagation},
  volume={62},
  number={6},
  pages={3242--3256},
  year={2014}
}

@techreport{3gpp38901,
   author={3GPP TR 38.901},
   title="{{Study on channel model for frequencies from 0.5 to 100 GHz}}",
   institution={3GPP},
   year={2023},
   type={Technical Report},
  
 }

@inproceedings{kang2024terrestrial,
  title="{Terrestrial-satellite spectrum sharing in the upper mid-band with interference nulling}",
  author={Kang, Seongjoon and Geraci, Giovanni and Mezzavilla, Marco and Rangan, Sundeep},
  booktitle={IEEE International Conference on Communications},
  pages={},
  year={2024}
}

@techreport{3GPP38821,
  author = {{3GPP TR 38.821}},
  title = "{Solutions for {NR} to support non-terrestrial networks ({NTN})}",
  institution = {3GPP},
  type = {Technical Report},
  year = {2023},
  month = mar,
  version = {V16.2.0},
  note = {Release 16}
}

@techreport{ITU-R-RS.2017-0,
  author = {{ITU-R RS.2017-0}},
  title = "{Performance and interference criteria for satellite passive remote sensing}",
  institution = {International Telecommunication Union, Radiocommunication Sector},
  type = {Recommendation},
  year = {2012},
}

@article{tsampazi2025system,
  title="{System-Level Experimental Evaluation of Reconfigurable Intelligent Surfaces for NextG Communication Systems}",
  author={Tsampazi, Maria and Melodia, Tommaso},
  journal={IEEE Transactions on Vehicular Technology},
  year={2025}
}

@article{tsiropoulou2015combined,
  title="{Combined power and rate allocation in self-optimized multi-service two-tier femtocell networks}",
  author={Tsiropoulou, Eirini Eleni and Vamvakas, Panagiotis and Katsinis, Georgios K and Papavassiliou, Symeon},
  journal={Computer Communications},
  volume={72},
  year={2015}
}

@inproceedings{wadaskar2025satellite,
  title="{Satellite-Terrestrial Coexistence in FR3 Band via Hybrid True-Time-Delay Array-based Nulling}",
  author={Wadaskar, Aditya and Cabric, Danijela},
  booktitle={IEEE International Symposium on Dynamic Spectrum Access Networks},
  pages={},
  organization={},
  year={2025}
}

@article{jain1984quantitative,
  title="{A quantitative measure of fairness and discrimination}",
  author={Jain, Rajendra K and Chiu, Dah-Ming W and Hawe, William R and others},
  journal={Eastern Research Laboratory, Digital Equipment Corporation, Hudson, MA},
  volume={21},
  number={1},
  pages={},
  year={1984}
}

@misc{MATLAB:2025,
  author = {MathWorks},
  title = "{MATLAB (R2025a)}",
  year = {2025},
  howpublished = {Available online at \url{https://www.mathworks.com}}
}

@inproceedings{jaeckel20225g,
  title="{A 5G-NR satellite extension for the QuaDRiGa channel model}",
  author={Jaeckel, Stephan and Raschkowski, Leszek and Thieley, Lars},
  booktitle={IEEE Joint European Conference on Networks and Communications \& 6G Summit},
  pages={},
  year={2022},
  organization={}
}

@article{li2002analytical,
  title="{An analytical model to predict the probability density function of elevation angles for LEO satellite systems}",
  author={Li, Sheng-Yi and Liu, CH},
  journal={IEEE Communications Letters},
  volume={6},
  number={4},
  pages={},
  year={2002},
  publisher={}
}

@book{balanis2016antenna,
  title="{Antenna theory: analysis and design}",
  author={Balanis, Constantine A},
  year={2016},
  publisher={John Wiley \& Sons}
}

@book{van2002optimum,
  title="{Optimum array processing: Part IV of detection, estimation, and modulation theory}",
  author={Van Trees, Harry L},
  year={2002},
  publisher={John Wiley \& Sons}
}

@article{bazzi2025upper,
  title="{Upper mid-band spectrum for 6G: Vision, opportunity and challenges}",
  author={Bazzi, Ahmad and Bomfin, Roberto and Mezzavilla, Marco and Rangan, Sundeep and Rappaport, Theodore and Chafii, Marwa},
  journal={arXiv preprint arXiv:2502.17914},
  year={2025}
}

@article{agarwal2023coexistence,
  title="{Coexistence assessment and interference mitigation for 5G and Fixed Satellite Stations in C-band in India}",
  author={Agarwal, Avinash},
  journal={arXiv preprint arXiv:2312.16079},
  year={2023}
}

@inproceedings{niloy2024ascent,
  title="{ASCENT: A Context-Aware Spectrum Coexistence Design and Implementation Toolset for Policymakers in Satellite Bands}",
  author={Niloy, Ta-Seen Reaz and Kumar, Saurav and Hore, Aniruddha and Hassan, Zoheb and Dietrich, Carl and Burger, Eric W and Reed, Jeffrey H and Shah, Vijay K},
  booktitle={International Symposium on Dynamic Spectrum Access Networks (DySPAN)},
  pages={240--248},
  year={2024},
  organization={IEEE}
}

@article{niloy2023interference,
  title="{Interference Analysis of Coexisting 5G Networks and NGSO FSS Receivers in the 12-GHz Band}",
  author={Niloy, Ta-Seen Reaz and Hassan, Zoheb and Stephenson, Nathan and Shah, Vijay K},
  journal={IEEE Wireless Communications Letters},
  volume={12},
  number={9},
  pages={1528--1532},
  year={2023},
  publisher={IEEE}
}

@article{wu2023space,
  title="{Space-Ground Multicast Group Control for Multiuser LEO Satellite Networks}",
  author={Wu, Dapeng and Qin, Chen and Cui, Yaping and He, Peng and Wang, Ruyan},
  journal={IEEE Transactions on Wireless Communications},
  year={2023},
  publisher={IEEE}
}

@article{ma2023resource,
  title="{Resource Scheduling for High-Capacity Multicast Service in Ultra-Dense LEO Satellite Networks}",
  author={Ma, Ting and Qian, Bo and Qin, Xiaohan and Zhang, Xin and Cai, Lin X and Zhou, Haibo},
  journal={IEEE Transactions on Vehicular Technology},
  year={2023},
  publisher={IEEE}
}

@article{heydarishahreza2024spectrum,
  title="{Spectrum sharing and interference management for 6G LEO satellite-terrestrial network integration}",
  author={Heydarishahreza, Navid and Han, Tao and Ansari, Nirwan},
  journal={Communications Surveys \& Tutorials},
  year={2024},
  publisher={IEEE}
}
\end{document}